\documentclass[10pt, book]{amsart}
\usepackage{amsmath}
\usepackage{latexsym}
\usepackage[all]{xy}
\usepackage{color}
\newtheorem{teorema}{Theorem}[section]
\newtheorem{definicion}[teorema]{Definition}
\include{amslatex}
\newtheorem{proposicion}[teorema]{Proposition}

\newtheorem{ejemplo}[teorema]{Example}

\usepackage[margin=1.2in]{geometry}

 \numberwithin{equation}{section}

\include{amsmath}
\begin{document}
\begin{title}
{\large{ Quantum Systems from Geometric Evolutions}}
\end{title}

\author{Ricardo Gallego Torrom\'e}

\maketitle
\begin{abstract}
In the framework of deterministic finslerian models, a mechanism
producing dissipative dynamics at the Planck scale is discussed.
It is based on a geometric evolution from Finsler to Riemann
structures defined on the fiber bundle ${ TM}\to
{ M}$. Quantum states are equivalence classes, composed by the
configurations that evolve through an internal dynamics drive by
the above geometric evolution. Each equivalence class is conformed
by the ontological states that evolve to the same final state. The
existence of an hermitian scalar product in an associated linear
space is discussed and related with the quantum pre-Hilbert 
space. This hermitian product emerges from geometric and
statistical considerations. Our scheme recovers the main ingredients of the usual Quantum Mechanics. 
Several consequences are discussed and compared with the predictions of the standard Quantum 
Mechanics. A natural solution of the cosmological constant
problems is proposed, as well as a mechanism for the absence of
quantum interferences at classical scales.
\end{abstract}

\tableofcontents{}

\section{Introduction}

\subsection{Motivation}

The aim of the present work is to introduce a consistent scheme capable to reproduce 
generic quantum systems as a result of an hypothetical, more basic
deterministic dynamics at the Planck scale.

The framework presented in this work is rather different than the
usual Quantum Theory: we suggest the possibility deterministic
systems at the Planck scale in such a way that do not delete the
role of Quantum Mechanics as a consistent theory at atomic,
molecular, nuclear and particle scales.
The objective was to recover all the main 
ingredients of the Quantum Theory and find testable consequences
for the new approach. We hope our results are enough to obtain
falsifiable tests of our ideas.

Any attempt to go beyond the actual state of the Quantum Theory should try to address typical questions. 
Quantum Mechanics works perfectly in their microscopic
applications (that is, atomic, nuclear, particle level, for
instance),
while local hidden variables theories are found 
problematic experimentally and the recurse to non-local variables, although logically possible, 
seems not really appealing or natural if Fundamental Physics is
local and the return to more understandable frameworks is the objective. Then, why should we search 
another theory, rival of the actual Quantum Theory? And should
this new theory be a Hidden Variable Theory?

Nevertheless the success of Quantum Mechanics, there are odd
questions that seems deep pathological problems of Quantum
Mechanics.
The existence of two different types of 
fundamental processes in Quantum Mechanics, namely, measurement and evolution processes, is
uncomfortable and apparently an intermediate state of the theory.
Another reason for a criticism of the usual Quantum Theory is the
permanent strong interpretational problematic of Quantum Mechanics
and the ontological character or not of its description.
Not only is that we can not make any space-time image for quantum processes, 
but that any causal, deterministic picture seems not working
naturally. The ambition of understanding in a geometric way seems absent in the orthodox doctrine and 
methods of the standard Quantum Theory.

Together with these subjective reasons, there are other problems
that are more real, in the heart of Quantum Mechanics:
\begin{enumerate}
\item Combining Quantum Theory with gravity seems wild elusive (at
least in dimension 4), nevertheless the strong attempts of
physicists along years. It could be because a fundamental key is
missed. Maybe is just the non-compatibility of Quantum Mechanics
in its actual state with gravity.

\item The cosmological constant problem is a key problem, again
with gravity as the non-quantum factor that makes non-consistent
the calculations performed until the present.

\item How is it possible that at large scales there are no quantum
interferences? Is it possible to find a local solution for it?

\item Quantum cosmology is simply impossible as a scientific
theory, because by the essence of the definition of what it is, is
not possible to disprove.
\end{enumerate}
However the above points, examining the formalism of Quantum Mechanics, one 
is moved to think that the orthodox interpretation of
the theory are the most natural ones. 
It seems there is a natural relation between them that makes any other attempt for interpreting 
Quantum Mechanics not so natural. If this is accepted, then Quantum Mechanics involves in a 
natural way its own problematic nature to be understood in a
realistic and geometric way.

This state of the art legitimates new perspectives. The natural is that the
new framework must be a pre-Quantum Theory 
because experience shows that we live in a world on which Quantum
Mechanics works at some scales. By pre-quantum theory we mean
that quantum mechanics is complete at the scale it is applicable.
However, in a pre-quantum description as the presented in this
work, it is not the most fundamental description at other scales.

The main idea of our approach is the following: we postulate the
existence of a hidden dynamics. This dynamics is induced by
geometric evolutions. We interpret that this happens along a
second compact time. The evolution of the fundamental degrees
under this dynamics induces the notion of quantum state. This fundamental 
dynamics is supposed to happens at the Planck scale.
Although being deterministic, this dynamics produces information 
loss, and this phenomenon is essential in the generation of
quantum states, which we define as equivalence classes.

Some of these are similar to ideas appeared originally in  $([1])$, where are investigated different 
examples of deterministic models and provides physical mechanisms
producing information loss,
using directly quantum mechanical tools. 
Nevertheless, our approach $([2])$ is based 
on a very different construction: loss information process happening when a (dual) Finsler structure 
on the bundle ${ T M}\to { M}$ evolves to
a Riemannian structure, also living on ${ T M }$ ([2]).
The topological space ${ M }$ (we will assume that is a smooth manifold) is the 
the configuration manifold of all the degrees of freedom of the physical system at the Planck 
scale. The basic mathematical constructions involving this average
or evolution
functor are developed in $[3]$, while some 
mathematical results used in this paper are presented in the {\it
Appendix A}.

In reference $[2]$ we have introduced our mechanism at the level of geometric 
structures, required to obtain bounded hamiltonian, but we did not
describe how this evolution generates
quantum non-local states. In the 
present paper we try to fill this gap. In addition, some new mathematical results and physical 
applications are included.

The general relation found between deterministic theories and a special construction from Randers 
spaces ({\it theorem $2.1$} and {\it theorem $2.2$}) is on the basis of our approach. This 
relation is general enough to accommodate in a geometric context any deterministic system capable 
to be formulated using Hilbert space theory, when some physical
requirements
hold (they are maximal speed and maximal acceleration). Indeed this 
connection can be taken as the logical justification for our
approach. It is a natural map, suggesting the mathematical
frame-work for a family of dynamical systems.

\subsection{Structure of the paper}

In {\it section $2$}, the basic elements and 
notations of deterministic finslerian systems are reviewed. We
provide a existence theorem that links quantum systems and
deterministic systems.

In {\it section $3$}, we introduce the 
principal notions of the Quantum Theory from deterministic
fisnlerian systems:
we present a notion of quantum state and after 
associating a ``vector" of a linear space, a separable
pre-Hilbert space with an hermitian scalar product and 
introduce a geometric description for quantum observables.
We draw the picture of a quantum measurement 
theory based on this geometric point of view.

In {\it Section $4$}, the concept of  two-dimensional time 
is motivated from the structure of the proof of mathematical results of  $[3]$. In order to 
understand the ``apparent" quantum correlations of EPR experiments, the notion of double event is 
introduced and related with the geometric formulation. We explain
the notion of double dynamics, in the basis of our  mechanism for
the
generation of the quantum states. A theoretically testable 
prediction is also given related with the limit of the quantum
correlations. Interference experiments and quantum correlations
are discussed.

In {\it Section $5$}, a scattering operator obtained from the transition amplitudes is introduced.
Some its properties like unitary property of the associated 
$S$-operator are proved. This operator, however is more general
than the usual one in Quantum Systems. Therefore we didn't solve
the problem of how to associate the operator with the Quantum
Mechanical operator.

In {\it Section $6$}, a short discussion of the contents is
presented relating some results presented in this paper with other
investigations.
Possible effective geometries approaches related with to our 
theory are presented. Our scheme is compared with the work of 't
Hooft on Deterministic Quantum Mechanics,
 remarking in this case the 
differences between both systems. In {\it Appendix A}, we recall the notions and results of 
Finsler geometry used in this work. Only proofs are presented for the new statements not found in 
the references. In {\it Appendix B} we present a dictionary between the elements appearing in 
Hamilton-Randers deterministic models and their equivalence in the quantum mechanics formalism. In addition, 
we collect the main predictions of our theory and compare them with the corresponding predictions of 
the Quantum Theory. Finally, we shortly discuss the relevance of
the different tests for our proposal.

\section{Deterministic Hamilton-Randers Models at the Planck scale}
\subsection{Notation and basic hypothesis}

Let ${ M}$ be the configuration manifold describing all the degrees of freedom at the 
Planck scale of a closed physical system. We want to define
evolution equations. In particular, the evolution equation are
first order differential equations. The flow is on the tangent
space of the configuration manifold { TM}. The tangent space
${ T}_x{ TM}$ and the dual space ${ T}^*_x{ TM}$ is
defined in a similar way. Gluing together these spaces we define
the bundles ${ TTM}$ and ${}^*{ TM}$ in the usual way.
The evolution of the deterministic system is determined giving a
initial point on ${ TM}$.

The theory presented in this work is based on
the following fundamental hypothesis, relating the ontological dynamics at the 
Planck scale with the existence of a microscopic time arrow:
\begin{enumerate}
\item There is a microscopic time arrow. It is associated with a
non-symmetric dynamics. This evolution takes place along an
internal and compact time $t$.

\item There is a Hamiltonian function associated with the time
inversion respect the time $s$, $I_s $. The hamiltonian function
has the property that generates an evolution operator such that it
is invariant under $I_s$.
\end{enumerate}
The reason why we adopt this radical principles we have to say
that, although the $H$-principle of Boltzmann  provides an
explanation for the second law of the thermodynamics, it is not
applicable to non-reproducible events, like the universe. The
universe is a whole irreversible because it is only one.

Since geometry offers the requirements for an objective
description of nature, we look for a natural geometric structure
able to contain an irreversible character. Non-reversible Finsler
structures appears as candidates.
The relation between Finsler structures and deterministic systems is based on the 
following hypothesis:
\begin{enumerate}
\item The ontological states at the Planck scale are described by points of the phase space ${ 
T^* TM}$; the tangent bundle ${ TM}$ is equipped with a dual Randers metric $F^*$ (Def 
A.2):
\begin{displaymath}
F^* :{ T^* T M}\to { R^+ }
\end{displaymath}
\begin{displaymath}
(x,p)\to \alpha(x,p)+\beta(x,p).
\end{displaymath}
The $\alpha$ term is a Riemannian metric; the $\beta$ term is
linear in momentum:
\begin{displaymath}
\alpha=\sqrt{a^{ij}(x)p_i p_j}, \quad\beta(x,p)=p_i b^i(x).
\end{displaymath}
The value of the form $b$ is bounded by the Riemannian metric
$\alpha$.

\item { Hypothesis on the ergodicity of the internal evolution}
The average on the phase sphere ${ S^*}_x{ TM}$ is
equivalent to the time-average along the internal time $t$ The
tangent sphere are defined using the corresponding Riemannian norm
$\alpha$.

\item { Hypothesis on the final equilibrium state of the
system} For large times $t\rightarrow T_{max}$, the physical
system tends to the equilibrium, defining the averaged state.

\item The reduction of the space of ontological states to the quantum mechanical Hilbert space is 
induced by the reduction of the Randers structure $({ TM},F^* )$ to the Riemannian 
structure $({ TM},h )$ defining the $U_t $-evolution. This
evolution can be written in the form
\begin{displaymath}
U_t :({ T M},F^* )\to ({ T M},g_t )
\end{displaymath}
\begin{displaymath}
g \to \frac{1}{T_{max}}((1-t)g_t +t h) ,\, t\in
[0,T_{max}],
\end{displaymath}
for a convenient choice of the time $t$.  The parameter $t$ labels
the evolution through the internal time. It is normalized to have
a maximal value $T_{max}$
\end{enumerate}

We postulate that the above evolution in the geometric structure
$F^* \rightarrow h$
corresponds to the average of the 
initial Finsler structure investigated in reference [3]. The
particular form of the evolution of the geometry is however, not
still prescribed.
\subsection{The Hamiltonian Function}

The Hamiltonian function is constructed in the following way. First, consider the Randers structure 
$({ T M},F^* ) $ with Randers function
\begin{displaymath}
F^*:{ T}^*{ TM}\to { R}
\end{displaymath}
\begin{displaymath}
 F^* (x,p)\to
\alpha(x,p)+\beta(x,p).
\end{displaymath}
This function is a special case of Cartan space (the difference
with usual Finsler case is that in the case of a Finsler space the
form $F^*$ lives on a tangent bundle and for Cartan spaces it
lives on a cotangent bundle. It is important that is positive
defined.

The Hamiltonian of a deterministic system is given by the function
\begin{displaymath}
{ H}:{ T}^*{ TM}\to { R}
\end{displaymath}
\begin{equation}
{ H}=\sum^{2N}_{i=1} p_{i}f^{i}({x})+G(x),
\end{equation}
where $G(x)$ is an arbitrary function. The Poisson equations for
the canonical variables are:
\begin{displaymath}
\frac{d p_i }{dt}=-\{{ H},p_i \}=p_j \{f^j (x),p_i\}+
\frac{\partial}{\partial x^i}G(x)=p_i H^{ij}(x)+G_i,\,\,
i,j=1,...,2N.
\end{displaymath}
\begin{displaymath}
\frac{d q_i }{dt}=-\{{ H},x^i \}=f^i,\,\, i=1,...,2N.
\end{displaymath}
The functions $G_i$ are arbitrary, making compatible the dynamics
with the generalized Legendre transformations:
\begin{displaymath}
p_i =p_i (x^j, f^j).
\end{displaymath}
The associated Randers space is
\begin{displaymath}
{ H}\to 2\sum^{2 N}_{i=1} {\beta}^{i}(x)y_{i}, \quad (y^1 ,..., y^{2N})\in { 
T^*} _{(x,y)}{ T M}.
\end{displaymath}
This Hamiltonian is the result of consider the Hamiltonian of a set of pairs 
of identical particles, one evolving forward on time $t$ and Hamiltonian function $F^* (x,y)$ and another 
identical particle backward on time with Hamiltonian $F^*(I_s
(x),I_s (y))$. If the manifold ${ M}$ has dimension $N$, then
the elementary infinitesimal evolution operator is of the form
\begin{displaymath}
U(x,p,x+dx,p+dp)={ Id}-\i \,dt \,F^*(x,p)-I_s(\i \,dt
\,F^*(x,p)).
\end{displaymath}
The operator $I_s$ is anti-unitary and act on the $1-form$ $dt$
changing the sign,
\begin{displaymath}
U(x,p,x+dx,p+dp)={ Id}-\i \,dt \,F^*(x,p)-\i \,dt
\,I_s(F^*(x,p)).
\end{displaymath}
We take the following distinction on the coordinates,
\begin{displaymath}
(x_1, x_2)\to (x_1, -x_2).
\end{displaymath}
Invariance under canonical quantization of the sistem $(x,p)$
implies
\begin{displaymath}
(p_1, p_2)\to (-p_1, p_2).
\end{displaymath}
This implies that the effective Hamiltonian is
\begin{displaymath}
{ H}(x,p)=F^* (x,p)-F^* (I_s (x),I_s (y)) =\alpha +\beta -\alpha + \beta =2\beta =2 \sum ^{2 N} 
_{i=1} \beta ^i y_i.
\end{displaymath}
Identifying term by term with the non-symmetric part of the Randers function, we 
obtain the relations
\begin{equation}
 2{\beta}^{i}(x,p)=f^{i}(x,p),\quad p_{i}=y_{i},\quad i=1,...,2 N,
\end{equation}
and the corresponding ordinary differential equations determining
the evolution on time $s$ are
\begin{equation}
f^{i}={\beta}^{i}=\frac{dx^{i}}{ds},\quad i=1,...,2 N.
\end{equation}
This is the basis for the relation between deterministic
Hamilton-Randers systems and Randers spaces described in ref. [2]: given
any Randers space, we can construct a deterministic system with bounded
generalized speed and acceleration using the geometric data contained in the 
Randers structure, except for the function $G(x)$ which remains arbitrary.
Conversely, given a deterministic system, it is possible to associated a 
Randers space, although it seems there is not a unique canonical
way to do it ([2]). There is in principle an arbitrariness on the
selection of the metric $\alpha$.

This equivalence between Hamilton-Randers models (formulated on
cotangent bundle) and deterministic system is general and it is
not requirement that $\alpha$ be flat. However, if the initial
data is a truly Finsler-Randers structure (formulated on a given
tangent bundle), the correspondence is done only for flat $\alpha$
metrics.

\subsection{Canonical quantization: Bounded Hamiltonian Operator}
In order to relate some of the notions of Quantum Mechanics from
the mathematics of finslerian deterministic systems ii is useful
to consider the canonical quantization in the following way: the
coordinates $(x^i,p^i)$ are promoted to be operators (derivations)
over the algebra of smooth functions defined on ${ TM}$,
$\mathcal{F}_{ TM}$:
\begin{displaymath}
\hat{X^i}\psi(x)=x^i\psi(x);\,\quad \hat{P_i}\psi(x)=-\imath\hbar
\frac{\partial \psi}{\partial x^{i}},\quad i=1,...,2N.
\end{displaymath}
This representation holds obviously the Dirac
quantization,
\begin{displaymath}
[\hat{X^i}, \hat{P_j}]_D=-\i \hbar \delta^i_j;
\end{displaymath}
$[\cdot]_D$ is the Dirac bracket. Although we are working with
deterministic systems on ${ M}$, these quantization rule will
help to define the dynamics in a Hilbert space formulation. This
will be helpful to obtain the link between quantum dynamics and
the ontological dynamics.

The beables operators are defined as the set of operators $\{X^i ,\, i=1,..., 2 N\}$ which 
commute between them $[X^i ,X^j ]_D=0$ for each bi-dimensional
value of the
parameters $(s,t)$ and that completely define the evolution along the internal time $t$. The 
associated canonical operators are $\{{ \hat{P}}^i ,\, i=1,...,2 N\}$ and also by definition $[\hat{P}^i 
,\hat{P^j}]_D=0$ on functions $\mathcal{F}_{{ TM}}$. This
quantization is canonical because
\begin{equation}
[\hat{X}_i ,\hat{P}_j ]_D=\imath \hbar \delta _{i j} .
\end{equation}
Therefore, canonical momentum are not beables. They are an example
of changeables. The usual representation of the momentum operators
is given by
\begin{displaymath}
\hat{P}_j =-\i \, \hbar \frac{\partial}{\partial x^j},\quad
j=1,...,2N.
\end{displaymath}
In presence of curvature (a Cartan-Berwald can have curvature,
even if the metric ${\alpha}$ is flat), canonical momentum
operators should be replaced by covariant derivatives, in our case
associated with Chern's connection. However, when the connection
coefficients live on the manifold ${ TM}$, the canonical
commutation relations $(2.4)$ are the same. Let us denote the
covariant derivative formally by $D _i =\partial _i +\Gamma _i
(X)$, because we work with Berwald spaces, the Chern's connection
leaves in the base manifold ${ M}$, defining an affine
connection on { TM}. If we associate this new operator with the quantum 
mechanical operator, then
\begin{displaymath}
[\hat{X}_j ,\hat{D} _i ]=[\hat{X}_j , -\imath \hbar \partial
_i]=\imath \hbar  \delta _{i j}.
\end{displaymath}
In addition, due to curvature, new commutation relations appear:
\begin{displaymath}
[\hat{D}_i,\hat{D}_j]=R_{ij},
\end{displaymath}
being $R_{ij}$ the components of the curvature endomorphism tensor
(because $[\partial_i,\partial_j]=0$). This is for the Chern
connection. Although we are restricted to the Chern connection,
the quantization procedure and results are also valid for other
connections like Cartan's connection. Using the Chern's connection
one has the result that Randers-Berwald spaces happens when
$d\beta=0$. Therefore, there is a direct relation between
deterministic systems and part of the de Rham cohomology of the
manifold { TM}.

Since the Riemannian metric $h$ is the average of the initial
Finsler structure fundamental tensor, $h=\langle g\rangle $ and because the
connection for Berwald spaces are the ``same" than the Levi-Civita
connection associated with the metric $h$, we can follow using the
canonical momentum operators and canonical quantization in
presence of curvature. This is another argument to consider the
sub-category of Berwald-Randers spaces as the most interesting
Finsler (or properly Cartan) spaces for our physical application.
Let us note that Berwald structures are also interesting because they hold
 a generalized Equivalence Principle; living the connection on ${ TM}$, 
through a coordinate change in ${ T M}$, we can put all the connection coefficients equal to 
zero at a given point, in a normal particular coordinate system.
Although the idea is formal, what we need to transform this affine
connection in an affine connection in a sub-manifold ${
M}\subset { TM}$ is an extra structure, such that it is
preserved by the connection. This structure, indeed, has also to
be consistent with the action of $I_s$ on { TM}.

A major difficulty in the quantization of Hamiltonian $(2.2.1)$ is that it is not bounded from 
below, due to the linearity in the momentum operator. A procedure to get a bounded Hamiltonian is to 
consider the averaged Hamiltonian on the sphere ${
S}^*_x\subset { T}^*_x{ TM}$ that formally we write like
\begin{displaymath}
\langle { H}\rangle :=\int _{{ S}^* _x} { {H}}(x,p)|{\psi}(x,p)| ^2
d^{2N-1}p.
\end{displaymath}
In this formal expression, the co-tangent sphere ${ S}^* _x
{ TM}\subset { T}^* _x ({ TM})$ is defined by
\begin{displaymath}
{ S}^* _x :=\{ p\in { T}^* _x { TM}\mid  \alpha (x,p)=1,
x\in { T M} \}.
\end{displaymath}
$|{\psi}(x,p)| ^2$ is a weight function on the sphere ${ S}^* _x $ and it is determined by 
the Berwald-Randers structure $({ TM}, F^*)$. Since the measure
function $\psi (x,p)$ has a non-trivial dependence on $x$ and $p$,
the possibility to obtain a bounded function. However, since we
are integrating in a sub-manifold of the phase-space, we have to
provide an interpretation of this operation to define a truly
hamiltonian.

This Hamiltonian function was introduced suggested by the properties of the 
average operation that associates to each Finsler structure $({ M},F)$
a Riemannian structure $({ M},h)$.  The 
way $\langle { H}\rangle $ acts producing the evolution of a function $f\in
\mathcal{F}_{{ T^*TM}}$ defines the averaged Hamiltonian:
\begin{equation}
\frac{\partial f}{\partial s} =\int _{{ S}^* _x} \{f,{
{H}}(x,p)\}\,|{\psi}(x,p)| ^2 \, d^{2N-1}p.
\end{equation}
$\{\cdot,\cdot\}$ is the Poisson bracket defined in ${ T^*TM}$.
There is an alternative definition of $\langle { H}\rangle $,
\begin{displaymath}
\frac{\partial f}{\partial s} =\int _{{ S}^* _x} \{f,{
{H}}(x,p)\,|{\psi}(x,p)| ^2 \}\, d^{2N-1}p.
\end{displaymath}
Both definitions coincide if $|{\psi}(x,p)| ^2 $ is a constant of
motion,  $\{{ {H}}(x,p),\,|{\psi}(x,p)| ^2 \}=0$. It is clear
the possibility of this kind of functions. For instance, a
particular interesting example of function $|{\psi}(x,p)| ^2$
could be
\begin{displaymath}
|{\psi}(x,p)| ^2 = Pol^2(x,p)e^{-H(x,p)^2};
\end{displaymath}
where $Pol^2(x,p)$ is the square of a real polynomial with complex
solutions. The main reason adopt the first formula instead of the
second one, is that in this case the ontological Hamiltonian
follows directly from the discussion of {\it subsection 2.2}. The
measure is therefore hopefully explained by statistical
considerations. For that, of course, we need the specific form of
the evolution of the geometry. In contraposition, the second
general definition of the averaged Hamiltonian implies a non-clear
definition of the ontological Hamiltonian, becoming a mesh and
outside our main line of argumentation. In addition, problems in
the quantization will arise if one consider the second option.

Expression (2.3.2) defines the averaged classical Hamiltonian.
The averaged Hamiltonian $\langle { H}\rangle $ defines the 
dynamics of an ``averaged" physical system. It also induce how is
the average quantum hamiltonian. In the definition of the averaged
classical Hamiltonian we have adopted the idea that the
Hamiltonian function is the generator of the classical evolution
along the time $t$, through a usual continuity equation. It is
important to note that we are using the hypothesis on ergodicity
for the internal dynamics. In this way, we are replacing average
on time by statistical average.

The averaged Hamiltonian $\langle { H}\rangle $ is not the complete Hamiltonian of the 
macroscopic system; gravitational Hamiltonian should be added to $\langle { H}\rangle $, 
producing a total null Hamiltonian on physical states. This is compatible with evolution 
$\mathcal{{ H}}_{total}(x,p,t)\to 0$, if the total Hamiltonian function is defined by ${ 
H}(x,p,t)=F_t (x,p )-F_t (I_s (x), I_{t} (p))$. Therefore, we have
another way to introduce gravitational interaction.

The averaged Hamiltonian function (2.3.2) has an associated quantum operator $\langle { \hat {H}}\rangle $. 
This operator is defined by
the action on arbitrary elements of the Hilbert space representing states of defined generalized 
coordinates:
\begin{equation}
\hat{\langle { H}\rangle }(\hat{X},\hat{P})\mid x\rangle :=\int _{{ S}^* _x} { 
\hat{H}}(\hat{X},\hat{P})|{\psi}(x,p)\mid ^2 |p\rangle d^{2N-1}p.
\end{equation}
The averaged quantum Hamiltonian operator $\langle {
\hat{H}}\rangle (\hat{X},\hat{P})$ is a linear operator on the Hilbert
space of states with defined by the beables $|x\rangle $. However, we use
the basis of defined ontological momentum $\{\, \mid p\rangle  \}$. This
expression is well defined mathematically. The set of momentum
states are normalize such that the Riemannian norm is 1:
$\hat{P}^i \mid p\rangle  =p^i \mid p\rangle  $ with $\alpha(x,p)=1$ and
$\hat{X^i} |x\rangle  =x^i |x\rangle $. Since the quantization of the
ontological simplectic coordinates is canonical, the expression
for the action of the Hamiltonian is given by
\begin{displaymath}
\hat{\langle { H}\rangle }(\hat{X},\hat{P})\mid x\rangle :=\int _{{ \tilde{S}}^* _x} { 
\hat{H}}(\hat{X},\hat{P})|{\psi}(x,p)\mid ^2
|p+G(x)\rangle d^{2N-1}\tilde{p}.
\end{displaymath}
The function $G(x)$ is the translation produced by the operators $\hat{X}^i $ on the 
momentum state $\mid p\rangle $, computable from the canonical commutation relations and the form of the 
operators $\beta ^i (\hat{X})$ that appear in the ontological
Hamiltonian.

Generalized Legendre transformations, relating momentum
coordinates with speed coordinates, should also be imposed.
Nevertheless, our approach are not altered by the imposition of
these constrains. Let us consider the following formulation of
generalized Legendre conditions as constrains on the ontological
Hamiltonian. Then, the ontological Hamiltonian for one degree of
freedom is given by
\begin{displaymath}
{ H}(x,p)=\alpha (x,p) +\beta (x,p) \to \tilde{{
H}}(x,p) +\lambda ^i (p_i -p_i (x^j,\beta^j)).
\end{displaymath}
We postulate invariance of the constrain under $I_s$. which
implies that the hamiltonian of each pair of particles remain the
same,
\begin{displaymath}
{ H}(x,p) +\lambda ^i (p_i -p_i (x^j))-{ H}(I_s (x),I_s (p))
-I_s (\lambda ^i ) I_s ((p_i -p_i (x^j,\beta^j)))=2\beta(x,p).
\end{displaymath}
The reason why we maintain invariant the constrain under $I_s$ is
that the definition of canonical momentum as a function of the
coordinates $(x_+,x_-)$ must be the same for both kinds of degrees
of freedom, the ones that evolve forward and backward on internal
time $t$.
\subsection{Deterministic Hamilton-Randers Models and Dynamical Systems}
In this {\it subsection} we obtain theorems on the relations between
both categories of objects. Is in this sense an existence result,
the quantum version of some of the results of {\it subsection 2.2}.

All the terms appearing in the Hamiltonian $(2.3.2)$ are bounded and positive definite because the 
functions $\{ \beta ^i,\, i=1,...2N \} $ are bounded and also because we are integrating only over the 
sphere ${ S}^* _x$, which is a compact manifold. Therefore we
obtain the following result:
\begin{teorema}
Let $({ TM},F^* )$ be a Randers space. Then there is a
deterministic system
with bounded generalized acceleration and speeds, whose averaged 
Hamiltonian operator is defined by the relation $(2.3.2$. The
average Hamiltonian is bounded.
\end{teorema}
$\langle { \hat{H}}\rangle $ is promoted to be the Quantum Hamiltonian
describing the evolution of the physical average systems (without
gravity), which we identify with a quantum system of general type.
\begin{teorema}
Let ${ \hat{H}}=2{\beta}^{i}(\hat{X})\hat{P}_{i}$ be a quantum Hamiltonian operator describing 
a deterministic system with bounded generalized accelerations and
speeds.
Then there is a Randers 
structure that reproduces the above Hamiltonian and the dual
Randers function is
\begin{displaymath}
F^*(x,y)=\sqrt{a _{ij}p^i p^j}+f_i (x) p^i .
\end{displaymath}
\end{teorema}
The Riemannian metric $a_{ij}$ is not defined from the original deterministic system. The 
criterion for it should be clarified when a dynamics for the
intrinsic Finsler geometry is provided.

These relations between models constructed from Randers spaces
 and dynamical systems motivate the use of Finsler models,
and in particular Randers spaces, in the construction of 
deterministic models at the Planck scale: it is a general map between two apparently different 
categories of objects which can be useful in the construction of consistent models of deterministic 
systems at the Planck scale and implies an intrinsic, microscopic time arrow. This microscopic 
time arrow is explicit because the non-symmetric property of the Randers metric. In addition, the 
half forward-backward construction resembles a kind of advanced-retarded solutions common in 
Quantum Electrodynamics, just formulated in an abstract, non-reversible
Hamilton-Randers phase space. This construction resembles the new ideas
about $E\rightarrow -E$ parity ([18]).

In our previous work ([2]) we did not obtain the quantization
rules and formalism corresponding to the usual Quantum Mechanics
from our proposal. This question is addressed in the following
{\it section}.

\section{Quantum Formalism from Geometric Evolution}
\subsection{Quantum States from Deterministic Hamilton-Randers Models}
In this {\it section} we show how the quantum formalism emerges
from deterministic finslerian models. For that, some basic
mathematical results from ref. [3] and {\it Appendix A} are used.
In particular, the main tool is an evolution in the tangent space
${ TM}$ induced from the geometric evolution $({
TM},F^*)\rightarrow ({ TM},h)$ of dual metric structures. This
will motivate a definition of non-local quantum states.

The center of mass of a convex body is the point $x$ that
minimizes the "total" distance function $d^2 _T (x,y)$, where $d_T
$ is the total Hamilton-Randers distance,
\begin{displaymath}
d^2_T(x)=\int_{{ K}} d^2 ((x,y),\xi) d\xi .
\end{displaymath}
 Let us consider that we start with
a convex body $\bar{K}\subset { TM}$.
Consider the transformations $\varphi _t$ producing the evolution of the left and right center of 
masses ({\it Theorem} $A.5$ in {\it Appendix A}),
\begin{displaymath}
\varphi _t :{ T M}\to { T M}
\end{displaymath}
\begin{displaymath}
m_r (0)\to m_r (t),
\end{displaymath}
\begin{displaymath}
 m_l (0)\to m_ l (t),
\end{displaymath}
where $m_r (t)$ and $m_l(t)$ are the right and left center of mass of the
convex and compact body $\bar{K}$, for the fundamental tensor $g_{t}$.
Then $m_1$, the center of mass for the Riemannian metric 
$h$, is a fixed point and indeed an attractor for $m_r (t)$ and $m_l (t)$. The whole set from $m_r 
(0)$ to $m_l (0)$ collapses to the point $m_1$ under this
evolution, induced from the $U_t$ evolution
(see {\it appendix A} for the notation and notions 
involved with this evolution). We denote the solutions of this
transformation $\varphi_t$ by the ``string" set $\gamma (t)$.

Given $x \in { T M}$, let us consider the maximal "string" produced by the above 
procedure of collapsing strings to a fixed point, expanding
maximally the initial
compact body $\bar{K}$ in such a way that the new string 
also collapses to $x$ in a finite time bounded by $T_{max}$. The
existence of the point is assured in Riemannian geometry; in our
case, Hamilton-Randers geometry, we do not have still a formal
proof, although it is very likely that also exists, under the same
hypothesis. Assuming that this is the case, we have two basic
possibilities:
\begin{enumerate}
\item The maximal set ${ K}$ is compact and smooth. Since the
generalized speeds of the ontological degrees of freedom are
finite, $T_{max}$, the time where the string is collapsed is also
finite. A technical difficulty with this argument that requires
for instance smoothness condition. Then ${ K}$ is compact.

\item That the maximal set ${ K}$ is not bounded, using the
ambient topology. Then, an arbitrary parameter is needed in order
to define a characteristic $T_{max}$ finite. Since the definition
of $T_{max}$ is linked with the geometric transformation, these
states are out from the formalism.
\end{enumerate}
Therefore we assume that  condition $1$ holds and it is assumed
that $\bar{K}$.

The attractor point during the geometric evolution $U_t$ is
invariant, because an isometry of $F _{t}$ is also an isometry of $h$ and $x$ is 
completely defined by the convex body ${\bar{K}}$ and by the
metric $h$ ({\it Proposition $A.6$}), that is also invariant.

Let us consider the set of all maximal strings constructed in this
way that have the same attractor point $x$. We denote this set by ${ K}_x$. Since the point $x$ is 
invariant through this $U_t$-evolution, it characterizes the quantum state. Indeed, to label the 
point $x$ we can use local coordinates in ${ T M}$, that we also by  $x$. If ${ K}_x$ is a 
sub-manifold of ${ TM}$, it can be locally described using coordinates, which we call normal 
coordinates $\{\phi _j, \,  j=1,...,dim\, ({ K}_x)\}$. These
coordinates can be extended
to form a local coordinate chart on ${ T M}$ containing the point $x$.
The complementary coordinates will be called 
co-normal $\{\pi _k\, , k=dim({ K}_x)+1,..., dim \, ({ T M})\}$ and their values are fixed 
for any point in ${ K}_x$,
\begin{displaymath}
\pi _k (z)=c_k (x),\, \forall z\in { K}_x.
\end{displaymath}
$c_k(z)$ are constant functions on ${ K}_x.$ ${ K}_x$ is
spread over $x\in { T M}$. This fact implies the non-local
character of the definition of quantum state that we will consider
later.

In order to characterize the quantum state 
${ K}_x$, all the coordinates of $x$ are not necessary. What characterizes it
is the value of the coordinates $\{\pi _k (x)\, , k=1,...,n-dim\, ({ K}_x)\}$ for the point 
$x$, because they do not change during the collapsing process
induced from the $U_t$ evolution.

Let us denote by $Inv_x :{ TM}\rightarrow { TM}$ the group
of transformations leaving invariant the point $x$, $Iso(F)$, the
group of isometries of the initial metric $F$ and by $Iso_x$ the
group of transformations, leaving invariant the point $x$ and
transforming maximal strings in maximal strings. Doing that, we
are assuming that the strings are in some sense, the same.
Starting with a maximal ${ K_0}$ with maximal string
$\gamma_{{ K}_0}$, the group $Iso_x$ is
\begin{displaymath}
Iso_x :=\{ g\in Inv_x\, \, | \,\, g\cdot \gamma_{{ K}_0}
=\gamma_{\gamma \cdot { K}_0} \}.
\end{displaymath}
With this notation, ${ K}_x$ is given by
\begin{displaymath}
{ K}_x =\{ \cup_{g}\, \gamma_{g{ K}_0},\, \, g\in Iso_x\}.
\end{displaymath}
This construction implies a quotient structure that is smooth.

The second ingredient determining the quantum state is the average
operation in momentum sphere, formally written as
\begin{displaymath}
\hat{O}|x\rangle =\int _{{ S^*}_x} |\psi (x,p)|^2 \hat{O} |p\rangle  dp.
\end{displaymath}

 The formal integral integration over the sphere ${ S}^* _x$ for any
operator is interpreted as the value after a finite evolution time $T_{max}$, when the 
system has evolved through every possible momentum $p\in {
S^*}_x { TM}$
such that the probability density to find the elementary system at $(x,p)$ is 
$|\psi (x,p)|^2$. In order to accomplish this in a finite time,
the system must have a finite extension, that we postulate
universal and of the Planck scale. The probability density is
defined in the following way,
\begin{displaymath}
|\psi (x,p)|^2=\frac{ dt_(x,p)}{T_{max}}.
\end{displaymath}
$dt(x,p)$ is the time the system needs to evolve from $(x,p)$ to
$(x+dx,p+dp)$ in the time $dt$.

The notion of quantum state that we have is dynamical and implies
that the quantum system is an open system. A comparison with a
physical system like a classical gas can clarify this point: while
the equilibrium state of a sub-system of the gas defines the
macroscopic static state, the microscopic state is always
dynamical, with interaction with the environment.

Both ingredients in the definition of the quantum state are
related because the ergodic character of the evolution on the
co-tangent space ${ S^* }_{x(s)} { TM}$. The evolution
generating the sub-manifold ${ K}_x$ is not independent of the
average operation. One of the relations is due to the existence of
transformations in ${ T^*TM}$ relating a subset of coordinate
$\{x_v \}$ of ${ TM}\subset{ T^*TM}$ with the canonical
ontological momentum $\{p_{v}\}$. These transformation are what we
denote as generalized Legendre transformations. The coordinates of
a generic point $x\in { TM}_x$ are $(\Phi, \pi).$ The
corresponding coordinates in the tangent space ${ TTM}$ are
$(\Phi,\pi,\dot{\Phi},\dot{\pi})$, where
$\dot{\Phi}=\frac{d\Phi}{ds}$, $\dot{\pi}=\frac{d\pi}{ds}$.
Formally, the $U_t$ evolution must be consistent with the Legendre
transformations. This formally holds if the tangent space has
locally the structure
\begin{displaymath}
{ T}_x{ TM}\cong { T}_{\Phi}{ K}\times {
T}_{\pi}{ L}.
\end{displaymath}
The associated co-tangent space is just defined like
\begin{displaymath}
{ T}^*_x{ TM}\cong { T}^*_{\Phi}{ K}\times {
T}^*_{\pi}{ L}.
\end{displaymath}
Since we have a Riemannian metric $h$ that is invariant under the
$U_t$-evolution, we can get a canonical transformation between
${ T}_x{ TM}\to { T}^*_x{ TM}$, which
conserve the structures,
\begin{displaymath}
{ T}_{\Phi}{ K}\to { T}^*_{\Phi}{ K},
\end{displaymath}
\begin{displaymath}
{ T}_{\Phi}{ K}\to { T}^*_{\Phi}{ K}.
\end{displaymath}
This duality is our definition of generalized Legendre
transformation.

The value of the coordinates that remain constant through the
geometric evolution characterize the quantum state. We can
understand these quantities in terms of symmetries of the initial
Finsler metric $F$ because the evolution $U_t$ is invariant under
isometries of the metric $F$. Therefore the set ${ K}_x$ admits
a modular group $G\subset SO(2N)$ that contains the group of
isometries of $Iso(F)$. The group $Iso_x$ is a possible antecessor
of gauge symmetry.
 \subsection{Topology and classification of fundamental systems. Composite systems}
The notion of the configuration manifold ${
M}$ associated with the ``universe" depends 
on the particular system. There is a minimal dimension, because
for dimension less than $2$, Berwald spaces are also Riemannian
spaces: our formalism is not applicable for $dim({ TM})\langle  3$. It
is natural to associate the minimal dimension manifolds presenting
dissipative geometric evolution with elementary quantum systems,
that we recall are couple of pair of particles. In this case, the
minimal dimension is $4$. Therefore, the classification of
elementary models is provide by the cohomology group $({ H}^1,
{ R})({ TM})$ of dimension $4$. Even more, we know that on
this ${ TM}$ there is a splitting, which implies ${ TM}={
TM_+}\times { TM_-}$. Kunneth formula gives the following,
\begin{displaymath}
{ H}^1 { TM} \cong { H}^1{ TM_+}\otimes { H}^0{
TM_-}\oplus{ H}^0{ TM_+}\otimes { H}^1{ TM_-}.
\end{displaymath}
Therefore, it is enough to calculate the first and zero cohomology
groups of $2$-dimensional tangent bundles. This is equivalent to
the cohomology ${ H}^1{ TM_+}={ H}^1\otimes { M_+}$.
Here, the cohomology of the $1$-dimensional manifold is known. For
${ M_+}$ and ${ M_-}$ compact, both must be homeomorphic to
the sphere ${ S}^1$, one has that the non-trivial cohomology
groups are given by ${ H}^1({ S}^1)\,\cong { H}^0\,({
S}^1)\,\cong{ R}$. Therefore, the fundamental systems are
defined by fields with values on ${ R}^4,$
\begin{displaymath}
{ H}^1({ TM})\,\cong { R}^2\oplus { R}^2.
\end{displaymath}
From the structure of the bundle and the cohomology, we can
introduce a complex structure,
\begin{displaymath}
{ H}^1({ TM})\,\cong { C}\oplus { C}.
\end{displaymath}
This representation of the value of the cohomology changes on the
time s only if the evolution on $s$ is non-continuous. The
simplest way is assume a discretization on time at this scale and
that the evolution is non-continuous. However, we can restore a
continuous description of the evolution on time $t$, if we
associate a curve in the vector bundle of spinors $(2,1)$. The
idea is that evolution on $s$ of a fundamental pair can be
described using a spinor along a submanifold of dimension $1$ on a
$(2,1)$-space-time. This is because the cohomology group of the
tangent bundle ${ TM}$ coincides with the Dirac spinors over a
flat manifold of dimension $3$.

Other systems can be described by almost cartesian product of these
fundamental manifolds (one procedure to obtain composed systems is
presented in [2]). There are at least two ways to produce a bigger
Randers space using just the above geometric data:
\begin{enumerate}
\item The first way is valid for complete general structures
\begin{displaymath}
\alpha =\alpha _1 \oplus \alpha _2 ;\,\, \beta =\beta _1 \oplus
\beta _2 .
\end{displaymath}
This construction does not produce interaction terms in the total Hamiltonian. There is a 
priori not relation ${\alpha}_1$ ${\alpha}_2$.
\item The second form recovers the impossibility for a external observer to differentiate  
between identical particles:
\begin{displaymath}
\vec{p}=\vec{p}_1 \times \vec{0} +\, \,\vec{0}\times \vec{p}_2 ;\vec{\beta}=\vec{\beta}_1 
\times \vec{0} +\, \,\vec{0}\times \vec{\beta}_2 ,
\end{displaymath}
\begin{displaymath}
\alpha =\alpha _1 \oplus \alpha _2;\, \, \alpha _1 =\alpha _2.
\end{displaymath}
The quantum total Hamiltonian is given by:
\begin{displaymath}
\vec{\beta}(\vec{p})=\big(\frac{1}{2}\vec{\beta} _1 (\vec{p}_1)+\vec{\beta} _1 
(\vec{p}_2)+\vec{\beta} _2 (\vec{p}_1)+\vec{\beta} _2
(\vec{p}_2)\big).
\end{displaymath}
The mixed terms produce the interaction. The condition $\alpha _1 =\alpha _2$ is to ensure 
that the above construction is a Randers space.
\end{enumerate}
The above discussion implies that there is a natural notion of
fundamental state. However, we can speak of fundamental state in a
closer way to Quantum Mechanics,
\begin{definicion}
Let us denote the sub-bundle ${ S^* } { K}_x:=\{{ S}^*_x { TM},\quad x\in 
{ K}_x \}\subset { T^* T M}$. This manifold defines the
fundamental quantum state $|x\rangle $.
\end{definicion}
{ remarks:}
\begin{enumerate}
\item We assume that the topological spaces (topology induced from
the ambient space by immersion) are manifolds, and indeed smooth
manifolds. It is a non-trivial thing and we do not have a proof for
it. However, we hope that this matter is only technical and even if
the sets are not strictly smooth manifolds, we are allow to do the
same kind of manipulations that we are doing, that consists
basically on have a notion of average of functions in the
corresponding topological space and that this average is geometrical
in nature, that is, independent of the description.

\item The notion of fundamental state works even for large
dimensions on the tangent bundle { TM}. Therefore, in order to
not get confuse, we denote the fundamental states of the spaces of
minimal dimension by fundamental minimal states. The main point can
be summarize in the following conjecture,
\paragraph{}
{\it Any fundamental state can be described as interacting minimal
states.}
\paragraph{}
\end{enumerate}
The kind of interaction could be an obvious generalization of the
interactions described before. From the relation between fundamental
minimal states and spinors in $3$-dimensional space, we conclude
that:
\begin{enumerate}
\item The fundamental minimal states at the Planck scale is described by an abstract quantum spin
system.

\item The dimension of the manifold where the spinors are defines
is three.

\item The system is not deterministic. Although we started with a
ontological dynamics that is completely deterministic, the
description in terms of spinors, that we call fundamental spinors,
is quantum mechanical, because we are mixing both types of
coordinates on $(\Phi,\pi)$ on  { TM}.
\end{enumerate}

\subsection{Associated Quantum Transition Amplitudes}
A fundamental quantum state is not a vector element of a Hilbert
space, but it has an associated vector in a linear space.
In order to show this, we introduce the amplitude 
transition for the evolution from the  state  ${ K}_p$ to the
fundamental state ${ K}_q$, where $p$ and $q$ are now points of
${ TM}$). Let us consider a fundamental degree of freedom, which
by construction contains a half-system evolving backwards  and an
identical half-system evolving forward on the internal time and
through the $U_t$ dynamics. The fundamental system is assumed
localized at the point $z\in { K}_p\cap { K}_q$ at instant
$t$. An inverted evolution from $z$ to $p$ makes possible to speak
of a evolution from $p$ to $q$, passing through the intermediate
point $z$.Recall that this evolution is associated to the collapsing
of the strings sets. Repeating the same 
procedure for any point of the intersection ${ K}_p\cap { K}_q$, because of the definition 
of the ontological evolution, we write down the value of the
transition amplitude for a transition between points $p$ and $q$
when the transition happens in a intermediate point $z\in {
K}_p\cap { K}_q$,
\begin{displaymath}
\int _{{ K}_p\cap { K}_q} e^{\imath \frac{1}{L}(d_F (p,z)+d_F (z,q)-(d_F 
(z,p)+d_F (q,z)))}dz.
\end{displaymath}
In the exponential function, we should take the distances in the
following way,
\begin{displaymath}
d_F (p,z)=inf\{ \int _{ {\gamma}(t)} \sqrt{g
(\dot{\gamma}(t),\gamma(t))},\,  \gamma :p \rightarrow z\}.
\end{displaymath}
$\gamma :p \rightarrow z$ is a continuous path joining $x$ with $z$.
The volume form $dz$ is the induced measure obtained from the
Hamilton-Randers volume form. From the standard volume on the tangent
space { TM}, which is a $2N$-form on the space ${ TM}$
\begin{displaymath}
w=\sqrt{det\,{g}}\, dy^1  \wedge \cdot \cdot \cdot \wedge dy^N.
\end{displaymath}
This is still not the required measure, since the set ${ K}_p\cap
{ K}_q$ is a sub-set of the space tangent bundle ${ TM}$. The
induced measure is what we call $dz$. Since we are use to integrate
on smooth manifolds, we will assume that this also the case,
although we remark again that we do not have a proof of this assert.

The above integral is not the most general transition amplitude
between fundamental states. We have to consider other possible
non-trivial paths between $p$ and $q$ than the direct intersection.
In particular, the most general transition from $p$ to $q$ via
fundamental states is given by the integral,
\begin{displaymath}
\int _{\cup_{b,c} { K}_c\cap { K}_b} e^{\imath \frac{1}{L}(d_F (p,z)+d_F (z,q)-(d_F 
(z,p)+d_F (q,z)))}dz_{cb};
\end{displaymath}
the elementary measure $dz_{bc}$ (that again we assume it is a
volume integral) is the induced measure in the corresponding
intersection ${{ K}_c\cap { K}_b}$In this integral the
fundamental states ${ K}_b,\, { K}_c$ are such that:
\begin{enumerate}
\item there are some $b, c$ such that ${ K}_b\cap \, {
K}_c\neq {\o}$,

\item the chain $\{{ K}_b\}$ links $p$ with $q$.

\item For different $a,b,c$, ${ K}_a\,\cap \, { K}_b\,\cap
{ K}_c= {\o}$,

\item Locality condition: since ontological states moves at a maximal Hamilton-Randers
speed, only chains that are contained in common region of local
influence are considered in the sum.
\end{enumerate}
With the above definition, the transition amplitudes between
fundamental states are coherent. That means that one sum over all
possible intermediate histories. Therefore, we define the
following transition amplitude,
\begin{definicion}
The transition amplitude from the fundamental state ${ K}_p$ to
the fundamental state ${ K}_q$ is defined by the chain
integral:
\begin{equation}
\langle p|q\rangle \,:= \sum_{chains} \int _{{\cup_{b,c} { K}_c\cap {
K}_b}} e^{\imath \frac{1}{L}(d_F (p,z)+d_F (z,q)-(d_F (z,p)+d_F
(q,z)))} dz_{bc},
\end{equation}
where $z\in { K}_p\cap { K}_q$ and the sum is over all the
possible chains connecting $p$ with $q$
\end{definicion}
The transition amplitudes are invariant under diffeomorphic
transformations in ${ S^* T M}$. This is relevant because the
degrees of freedom at the Planck scale are not identified with the
degrees of freedom to systems where Quantum Mechanics is applied.
The geometric origin of the transition amplitudes induce a
coordinate-free definition of quantum state, even if we use
explicit coordinates in the above definition. The transition
amplitudes could be interpreted as the fundamental objects to
investigate.
\subsection{Properties of the transition amplitudes. Quantum pre-Hilbert Space}
We check that the transition amplitudes $(3.2.1)$ has the expected
properties associated with the quantum symmetries.
The first one is linearity of the associated hermitian product. Let us define the 
transition amplitudes between an arbitrary fundamental state ${
K}_q$ and two orthogonal and fundamental states ${ K}_1 $ and $
{ K}_2 $ corresponding to the points $q_1$ and $q_2$. For each
term of the corresponding possible chain of the integrations we have
that for non-intersection ${ K}_1$ and ${ K}_2$ that
\begin{displaymath}
 \langle p|{ K}_1 \cup { K}_2 \rangle \,=\int _{{ K}_p\cap ({ K}_1\cup { K}_2)} dz \,e^{\imath 
\frac{1}{L}(d_F (p,z)-d_F (z,{ K}_1\cup { K}_2)-(d_F (z,p)+d_F ({ K}_1\cup 
{ K}_2,z)))}=
\end{displaymath}
\begin{displaymath}
:=\int _{{ K}_p\cap { K}_1}dz\, e^{\imath \frac{1}{L}(d_F (p,z)+d_F (z,q_1 )-(d_F (z,p)+d_F 
(q_1 ,z)))}+
\end{displaymath}
\begin{displaymath}
+\int _{{ K}_p\cap { K}_2} dz\,e^{\imath \frac{1}{L}(d_F (p,z)+d_F (z,q_2 )-(d_F (z,p)+d_F 
(q_2 ,z)))}.
\end{displaymath}
Here $dz$ denotes generically the corresponding measure. This result
come from the definition of the integration of chains. Then for
fundamental states ${ K}_{p}$, ${ K}_{q_1}$ and ${
K}_{q_2}$ that are orthogonal in the sense that the transition
amplitude are zero, the following equality holds:
\begin{equation}
\langle p|q_1 \cup q_2 \rangle \,= \langle p|q_1\rangle \, +\,\langle p|q_2 \rangle ,
\end{equation}
for arbitrary state $|p\rangle $. As a consequence, it is natural to define
the symbol $|q_1\rangle \,+\, |q_2\rangle $ as an anti-linear functional on the
dual space of formal linear combinations of fundamental spaces:
\begin{displaymath}
(\langle p|\,+\lambda\,\langle q|)|(|q_1\rangle \,+\, |q_2\rangle )=\langle p|(|q_1\rangle \,+\, |q_2\rangle )
\,+\,\lambda^*\,\langle q|(|q_1\rangle \,+\, |q_2\rangle ).
\end{displaymath}
The linear combinations of bra are only formal, while the linear
combinations of kets are defined by the transition amplitudes
$(3.4.1)$. Therefore $(|q_1\rangle \,+\, |q_2\rangle )$ is an element that we can
associated with a vector either, producing the transition amplitudes
(3.4.1). In this way, an internal sum of fundamental states is
defined and associated amplitudes by $3.4.1$ are given.

Linearity by a complex scalar multiplication of the transition
amplitudes (3.3.1) is realized in the following way. First we use
the properties of integration on chains to see that for each term of
each possible change linking the states ${ K}_p$ and ${ K}_q$
one obtains
\begin{displaymath}
\langle p|\lambda q \rangle \,=  \int _{({ K}_p\cap \lambda { K}_q)}
e^{\imath \frac{1}{L}(d_F (p,z)+d_F (z,q)-(d_F (z,p)+d_F (q,z)))}.
\end{displaymath}
From the formal integration of changes the value of the integral is
defined usually as
 \begin{displaymath}
\langle p|\lambda q \rangle \,=\lambda  \int _{{ K}_p\cap { K}_q} e^{\imath \frac{1}{L}(d_F (p,z)+d_F 
(z,q)-(d_F (z,p)+d_F (q,z)))}.
\end{displaymath}
This relation defines the quantum state $\lambda { K}_q$ as the one producing the above 
transition amplitudes to fundamental states:
\begin{displaymath}
\langle p|\lambda q \rangle \,:=\lambda\,\langle p|q \rangle ,\,\, \forall \lambda\in {
C},\,\,\forall |p\rangle .
\end{displaymath}
We associate the vector $\lambda |q\rangle $ to the quantum state 
$\lambda { K}_q$.

$|q_1\rangle \,+\, |q_2\rangle $ is an element of the linear envelope generated by
the set of fundamental states $\{ { K}_q,\, q\in { TM}\}$. We
promote $|q_1\rangle \,+\, |q_2\rangle $ to
be a ``phenomenological quantum state". The 
difference between fundamental quantum sate and ``phenomenological quantum state" is that
fundamental quantum states are chains of order $n=1$, while the fundamental quantum states are 
larger chains. Therefore, fundamental states are the ``simplices",
defined as the set $\{{ K}_q,\,\, q\in { T M}\}$. From this
simplices one can define chains. The transition amplitudes
produces a morphism between these topological sets and the
category of pre-Hilbert spaces. The simplices are determined by
the $U_t$ evolution. It could be interesting to know if
corresponds to a known mathematical structure or if it has a deep
meaning.

One can define a vector space structure generated by  $\{{ K}_q,\, q\in 
{ T M}\}$. This linear space is endowed with a scalar product
with physical meaning. We should check the
 properties of this product for fundamental states.
 We show 
that it is indeed an hermitian scalar product. From the definition of the exponential function, it 
follows that
\begin{displaymath}
\langle p|q \rangle \,=\int _{{ K}_p\cap { K}_q} e^{\imath \frac{1}{L}(d_F (p,z)+d_F (z,q)-(d_F 
(z,p)+d_F (q,z)))}=
\end{displaymath}
 \begin{equation}
 \int _{{ K}_p\cap { K}_q} e^{-\imath \frac{1}{L}(d_F (q,z)+d_F (z,p)-(d_F (z,q)+d_F 
(p,z)))}=\langle q|p \rangle ^* .
\end{equation}
For phenomenological quantum states hermiticity is obtained through
the decomposition in terms of fundamental states and using
linearity.

The fundamental states is a pre-Hilbert space. It is a separable, by
construction. The question of the completeness of this pre-Hilbert
space is a problem on the convergence of manifolds. It requires
methods from modern geometry and metric theory.

Calculating the transition amplitude from one state into itself,
we obtain the condition for compact spaces
\begin{equation}
\langle q|q\rangle \, =\int _{{ K}_q} 1\,dz:= Vol({ K}_q)
\end{equation}
for an arbitrary fundamental quantum states ${ K}_q$.
In order to avoid any problem with divergences 
in the integration we should take compact domains of integrations,
corresponding to compact quantum spaces. Compact, fundamental quantum
states live in the projective Hilbert space 
because we can multiply by $1/\sqrt{Vol({ K}_q)}$ for compact or
bounded states ${ K}_q$. Therefore we can normalize in the
following way:
\begin{displaymath}
|q\rangle \, \to \frac{1}{\sqrt{vol({ K}_q})}|q\rangle .
\end{displaymath}
In case of non-compact states, such that we need $T_{max}\rightarrow
\infty$ to recover the whole state, we use the following
normalization:
\begin{displaymath}
|q\rangle \, \to \lim_{R\rightarrow
\infty}\frac{1}{\sqrt{vol({ K}_q(R)})}|q\rangle _R;
\end{displaymath}
$R$ indicates that we are only taking the intersection of the
quantum state ${ K}_q$ with the Riemannian ball of radius $R$ in
${ S^*} _q { TM}$ centered at $q$. We use the hypothesis we
make now is that we work with normalized states,
$\frac{1}{\sqrt{vol({ K}_q(R)})}|q\rangle _R$, then we perform
calculations involving homogeneous quantities of degree zero in $R$
(quotients of products of normalized vectors). Therefore, for large
$R$, we expect these calculations are unsensible to $R$, because
they are all homogeneous of degree zero in $R$..

Consider a basis of the pre-Hilbert,
separable space generated by all the fundamental, 
orthonormal states with null intersection $\langle { \psi^{k_i }} | {
\psi^{k_j}} \rangle = 0,\, \forall k_i \neq \forall k_j$,
\begin{displaymath}
\Xi :=\{\psi\,|\,\, \langle \psi^k | \psi^j \rangle =0,\,j\neq k\},
\end{displaymath}
\begin{displaymath}
\mathcal{H}:=\langle  \Xi \rangle _{ C};
\end{displaymath}
$\langle  \Xi \rangle _{ C}$ denotes the complex linear enveloping of $\Xi$.
Orthogonality does not mean that the specific states has non-trivial
intersection, but the set of chains that can produce this kind of
transition interfere in a destructive way. In general $\mathcal{H}$
is an infinite dimensional lineal space.

We want to check that the following identity holds for the
orthonormal states of $\langle  \Xi
\rangle _{ C}$,
\begin{equation}
I=\int_{{\Xi}} d\mu (\psi) |\psi\rangle \langle \psi|\,.
\end{equation}
$d\mu (\psi)$ is a convenient measure,
\begin{displaymath}
d\mu (\psi)=\Theta (\psi) d^k \psi
\end{displaymath}
and $\Theta (\psi)$ is the density distribution in the topological
set $\Xi$. Usually this measure is not smooth.

Let us consider two arbitrary states ${ K}_p$ and
${ K}_q$. Because the domain of intersections are empty, we immediately have a decomposition of the 
integration domain ${ K}_p\cap { K}_q$ as union of disjoint sets $\psi $ such 
that ${ K}_p\cap { K}_q=\cup _{k }\Psi^k $ with 
$\langle \Psi^{k_1}|\Psi^{k_2}\rangle \, =0$, for each possible chain joining the
states ${ K}_p$ and ${ K}_q$ and for each term of these
changes, we check that
\begin{displaymath}
\langle p|q\rangle \, =\int _{{ K}_p\cap { K}_q} e^{\imath \frac{1}{L}(d_F (p,z)+d_F (z,q)-(d_F 
(z,p)+d_F (q,z)))}\,=
\end{displaymath}
\begin{displaymath}
\sum _{k}\int _{\psi^k }e^{\imath \frac{1}{L}(d_F (p,z)+d_F (z,q)-(d_F (z,p)+d_F 
(q,z)))}\, =\int _{\Xi} d\mu({\psi}) \langle p|\psi\rangle \langle \psi|q\rangle \,.
\end{displaymath}
Since this relation holds for any intersection of any possible chain
for any pair of fundamental states, it holds by linearity for any
linear combination of them.
\subsection{Evolution of the Quantum Transition Amplitudes through the External Time s}
Let us calculate the value of the transition amplitude (3.3.1)
between two states at different instants $\langle q_{0}(s=0)|q_{n}(s=n)\rangle $
due to the evolution with the Hamiltonian (2.3.3) for quantum
states. The hamiltonian producing these transitions is the average
Hamiltonian. Using the decomposition of the unity operator (3.4.4),
the transition amplitude can be expressed as
\begin{displaymath}
\langle q_{0}(s=0)|q_{n}(s=n)\rangle =\langle q_{0}(s=0)\int_{\Xi} d\mu (q_{1})
|q_{1}(s_1)\rangle \langle q_{1}(s_1)
\end{displaymath}
\begin{displaymath}
\int_{{\Xi}} d\mu (q_2) |q_{2}(s_2)\rangle \cdot \cdot \cdot \int_{{\Xi}}
d\mu (q_{n-1})|q_{{n-1}}(s_{n-1})\rangle  \langle q_{n-1}(s_{n-1)}|q_{n}(s_n)\rangle .
\end{displaymath}
This transition amplitude is different that the transition amplitude
defining the ontological quantum states (3.1.1), because each
individual factor
\begin{displaymath}
\langle q_{j-1}(s_{j-1})|q_{j}(s_j)\rangle ,\, j=0,...,n-1
\end{displaymath}
is obtained using the average Hamiltonian (2.5). We promote this
element to be an usual quantum mechanical transition amplitude due
to an evolution. It is convenient to write the transition
amplitude as
\begin{equation}
\langle q_{0}(s=0)|q_{n}(s_n)\rangle =\prod^{n-1}_{j=2} \int_{{\Xi}}d\mu
(q_{1})\langle q_{j-1}(s_{j-1})|q_{j}(s_j)\rangle .
\end{equation}
The evaluation of the elements is just given by the expression
\begin{displaymath}
\langle q_{j-1}(s_{j-1})|q_{j}(s_j)\rangle  =\langle q_{j-1}(s_{j-1})|\langle { \hat{
U}}\rangle |q_{j-1}(s_{j-1})\rangle .
\end{displaymath}
We can compute this formula explicitly
\begin{displaymath}
\langle q_{j-1}(s_{j-1})|\langle { \hat{
U}}\rangle |q_{j-1}(s_{j-1})\rangle =\langle q_{j-1}(s_{j-1})|\big(I-\,\imath\,ds\,\langle {
\hat{ H}}\rangle (\hat{{ X}},\hat{{ P}})\big)|q_{j-1}(s_{j-1})\rangle =
\end{displaymath}
\begin{displaymath}
=1-\imath\,ds\,\langle q_{j-1}(s_{j-1})|\langle { \hat{ H}}(\hat{{
X}},\hat{{ P}})\rangle |q_{j-1}(s_{j-1})\rangle =
\end{displaymath}
\begin{displaymath}
=1-\imath\,ds\,\langle q_{j-1}(s_{j-1})|\,\int_{{
S}^*_{q(s_{j-1))}}}{ \hat{ H}}({{ X}},\hat{{
P}})|p\rangle \,|\psi(q_{j-1}(s_{j-1}),p)|^2 dp_{j-1}.
\end{displaymath}
Since we know the form of the ontological Hamiltonian ${ \hat{
H}}({{ X}},\hat{{ P}})$ we can following the computation,
\begin{displaymath}
\langle q_{j-1}(s_{j-1})|q_{j}(s_j)\rangle  =1-\imath\,ds\,\int_{{
S}^*_{q(s_{j-1))}}}|\psi(q_{j-1}(s_{j-1}),p)|^2\,\langle q_{j-1}(s_{j-1})|\,{
\hat{ H}}({{ X}},\hat{{ P}})|p\rangle \, dp_{j-1}=
\end{displaymath}
\begin{displaymath}
=1-\imath\,\int_{{
S}^*_{q(s_{j-1))}}}|\psi(q_{j-1}(s_{j-1}),p)|^2\,2\beta^i
(q_{j-1}(s_{j-1}))p_i\,\langle q_{j-1}(s_{j-1})|p\rangle \, dp_{j-1}.
\end{displaymath}
We need to compute the value of the product $\langle q_{j-1}(s_{j-1})|p\rangle $
and relate this value with the quantum mechanical amplitude. The
state $|q_{j-1}(s_{j-1})\rangle $ correspond to a quantum state, designed
by the coordinates $q_{j-1}(s_{j-1})$, which does not change during
the $U_t$-evolution. However, the value of the product
$\langle q_{j-1}(s_{j-1})|p\rangle $ are defining using the $U_t$ evolution
through a transition of the type (3.3.1). The state $|p\rangle $ is defined
using the Fourier transform
\begin{displaymath}
|p\rangle =\int_{{ TM}} \chi_p(x)\,|x\rangle \,dx.
\end{displaymath}
$\chi_p(x)$ is a solution to the Laplacian on the manifold ${
M}$ for the averaged Riemannian metric $h$ with eigenvalue $p^2$;
the integration is extended to the set of functions that can be
developed by eigenvalues of the Laplacian. Then, the product is
equal to
\begin{displaymath}
\langle q_{j-1}(s_{j-1})|p\rangle =\int_{{ TM}}
\chi_p(x)\,\langle q_{j-1}(s_{j-1})|x\rangle \,dx.
\end{displaymath}
Note that this is in general not a delta function because the states
${ K}_{q(s_j)}$ are in general not independent.

The amplitude transition is given by the formula
\begin{displaymath}
\langle q_{j-1}(s_{j-1})|q_{j}(s_j)\rangle  =1-\imath\,ds\,\int_{{
S}^*_{q(s_{j-1})}}|\psi(q_{j-1}(s_{j-1}),p)|^2\,2\beta^i(q_{j-1}(s_{j-1}))p_i\,\cdot
\end{displaymath}
\begin{displaymath}
\cdot\Big(\int_{{ TM}} \chi_p(x)\,\big(\sum_{chains} \int _{ {
K}_{q(s_{j-1})}\cap { K}_x} e^{\imath \frac{1}{L}(d_F
(q(s_{j-1}),z)+d_F (z,x)-(d_F (z,q(s_{j-1}))+d_F (x,z)))} dz\,\big)
\,dx\,\Big) dp_{j-1}.
\end{displaymath}
The sum of the intersections that appear and the chains must be
understood as the sum on the possible non-trivial chains joining the
initial and final states presented in {\it subsection} 3.4.

At first sight this equation seems independent of the last point.
However, this is not really the case, because due to the Legendre
transformations,
\begin{displaymath}
p_i (s_{j-1}) = p_i \big(s_{j-1}) (q_{j}(s_{j})-q_{j-1}(s_{j-1})),
q, p, \lambda\big);
\end{displaymath}
$\lambda$ are parameters of the system (like masses or other
constants of the evolution). Therefore, the Dirac representation of
a elementary quantum transition amplitude is given by the
expression,
\begin{displaymath}
e^{\imath S/\hbar }\sim \langle q_{j-1}(s_{j-1})|q_{j}(s_j)\rangle
=1-\imath\,ds\,\int_{{
S}^*_{q(s_{j-1})}}e^{log\Big(|\psi(q_{j-1}(s_{j-1}),p_{j-1}
\big(s_{j-1}) (q_{j}(s_{j})-q_{j-1}(s_{j-1})), q, p,
\lambda\big))|^2\Big)}\,\cdot
\end{displaymath}
\begin{displaymath}
\cdot e^{log\Big(2\beta^i(q_{j-1}(s_{j-1}))p_{i} \big(s_{j-1})
(q_{j}(s_{j})-q_{j-1}(s_{j-1})), q, p, \lambda\big)\Big)}\,\cdot
\end{displaymath}
\begin{equation}
\cdot\Big(\int_{{ TM}} e^{log \, \chi_p(x)}\,\big(\sum_{chains}
\int _{ { K}_{q(s_{j-1})}\cap { K}_x} e^{\imath
\frac{1}{L}(d_F (q(s_{j-1}),z)+d_F (z,x)-(d_F (z,q(s_{j-1}))+d_F
(x,z)))} dz\,\big) \,dx\,\Big) dp_{j-1}.
\end{equation}
Comparing terms, one get an infinitesimal action that is
\begin{displaymath}
{\imath dS/\hbar }\sim \imath\,ds\,\int_{{
S}^*_{q(s_{j-1})}}e^{log\Big(|\psi(q_{j-1}(s_{j-1}),p_{j-1}
\big(s_{j-1}) (q_{j}(s_{j})-q_{j-1}(s_{j-1})), q, p,
\lambda\big))|^2\Big)}\,\cdot
\end{displaymath}
\begin{displaymath}
\cdot e^{log\Big(2\beta^i(q_{j-1}(s_{j-1}))p_{i} \big(s_{j-1})
(q_{j}(s_{j})-q_{j-1}(s_{j-1})), q, p, \lambda\big)\Big)}\,\cdot
\end{displaymath}
\begin{equation}
\cdot\Big(\int_{{ TM}} e^{log \, \chi_p(x)}\,\big(\sum_{chains}
\int _{ { K}_{q(s_{j-1})}\cap { K}_x} e^{\imath
\frac{1}{L}(d_F (q(s_{j-1}),z)+d_F (z,x)-(d_F (z,q(s_{j-1}))+d_F
(x,z)))} dz\,\big) \,dx\,\Big) dp_{j-1}.
\end{equation}
From this definition, one gets a reality criterion in the
definition of the logarithm functions.

The transition function (3.5.2) is a genuine quantum mechanical
amplitude transition. Since the only label that is not integrated is
on the state $q_{j-1}(s_{j-1})$, the evolution is generally governed
by the $Schr\ddot{o}dinger $ equation. if $s_j -s_{j-1} =ds$, the
unitary operator is $\langle { \hat{U}}\rangle ={ I}-\frac{\imath ds}{\hbar}
\langle { \hat{H}}\rangle $ and therefore,
\begin{displaymath}
-\imath \hbar \frac{\partial}{\partial s}  |q(s)\rangle =\langle { \hat{H}}\rangle
|q(s)\rangle .
\end{displaymath}
\subsection{A Mechanism producing the absence of Interferences at large scales. The Classical Limit}
The distance $L$ in the equation $(3.3.1)$ is associated with a
universal length scale, different than the Planck scale. $L$ gives a
measure of the scales where interference phenomena can happen and
where they are absent.  Apart from the way that different chains
contribute to possible interferences, the value of $L$  provides the
scale of the quantum interferences. We will assume a universal value
for $L$.

In the case of a Quantum Field Theory with particles of different
species, one natural generalization is to consider instead of the
single numerical constant $1/L$ the inverse of a ``scale matrix"
 and consider an exponential functions of the form:
\begin{displaymath}
e^{\imath (d_{F M} (p,z)+d_{F M} (z,q)-(d_{F M} (z,p)+d_{F M}
(q,z)))},
\end{displaymath}
where the distances are obtained replacing the fundamental tensor
by:
\begin{displaymath}
F^* \to {F^* _M},
\end{displaymath}
that is a matrix-valued function. The corresponding fundamental
tensor is given by
\begin{displaymath}
g\to  g_M.
\end{displaymath}
This is a generalized metric structure, that we can denote matrix-valued metrics and denote
 $g_M\,\in\,F^* M $. Here $M $ stand for  ``matrix". Since the
addition of matrices is abelian, this generalized distance is
symmetric iff the matrix $M$ is symmetric as well. We require that
the matrix $M$ has all the coefficients positives. It is not
necessary to require non-degeneracy.

An associated idea is introduced in the theory of Modified
Dispersion Relations ([19]) in the context of Finsler space-time
geometries. In this theory, the dispersion relation that gives the
frequency as a function of the moment for elementary particle,
depends on the mass of the particles. This is associated with the
Hamilton-Randers character of the space-time metric.

Let us assume the simpler case of a constant and scalar value for
$L$. A decoupling for a long Finsler distance $d _F $ can happens,
because the integration of a highly oscillating function could be
zero. This corresponds with a large Riemannian distance ``$d_h$" in ${ TM}$, 
due to {\it Proposition A.7}. If this happens for any point of the intersection, there is a 
complete decoupling between the states ${ K}_p$ and ${ K}_q$
 (note that both $d_{F}$ and $d_R$ are distances 
in { TM}). Absence of quantum interferences is related with
orthogonality condition of states,
\begin{displaymath}
\langle p|q\rangle \,:= \sum_{chains} \int _{{\cup_{b,c} { K}_c\cap { K}_b}}
e^{\imath \frac{1}{L}(d_F (p,z)+d_F (z,q)-(d_F (z,p)+d_F (q,z)))}
dz_{bc}=0.
\end{displaymath}
This condition does not mean that ${ K}_p\cap { K}_q=\emptyset$. Indeed it can happens for short chains
 but that have non-zero intersection, due to a fast oscillating behavior of the exponential function
on the domain ${ K}_p\cap { K}_q$, the integral can be zero or
very small. The use of complex amplitudes is therefore justified by
this effect, that is a interference effect. For long chains,
different contributions can simultaneously act as eliminating the
total contribution, because of the re-distribution of the actual
ontological state in a bigger number of possibilities, reducing a
given transition to relative probability zero.

One geometrical reason why the transition amplitude can go to zero
is a large separation between states $p$ and $q$ (by a large distance we mean a large value of the 
exponent because one of the distances involved appears large
compared with the others). This property provides a mechanism to
understand the absence of quantum interferences at large scales. If
we re-write the exponential function as
\begin{displaymath}
e^{\imath \frac{1}{L}(d_F (p,z)+d_F (z,q)-(d_F (z,p)+d_F (q,z)))}
=e^{\imath \frac{1}{L}(d_F (p,z)+d_F (q,z)-(d_F (z,p)+d_F (z,q)))}
\end{displaymath}
the decoupling between physical systems happens
when any of the following conditions hold:
\begin{enumerate}
\item  A large difference between the forward distance and
the backward distance compared with the scale distance
$L$:
\begin{displaymath}
d_F (p,z)+d_F (q,z)\gg d_F (z,p)+d_F (z,q) + L.
\end{displaymath}
 It can be shown from some examples ([$5$]) that in Finsler geometry a large left distance $d_F 
(p,z)$ can be associated with a short right distance $d_F (z,p)$.
This asymmetry can be interpreted physically as a decoupling
associated with a irreversible evolution from the state ${ K}_p$
to the state ${ K}_q$. Alternatively, if we associate a material
point moving along a trajectory joining $z$ and $p$, one can also
interpret this condition as equivalent that the system has a large
energy, that is usually a characteristic of macroscopic objects.

\item The transition is produced between a ``relative local" state and a ``relative spreaded" state. 
Mathematically this situation can be described as
\begin{displaymath}
d_F (p,z)-d_F (z,p)\gg d_F (z,q)-d_F (q,z).
\end{displaymath}
This happens if all the points $z\in { K}_p\cap { K}_q$ are relative close to the point 
$q$ but relative far from $p$. The meaning of it is just that the possible evolutions from $p$ to 
$q$ are forbidden because one of the states has an associated energy  much large compared with
the other. This kind of decoupling also incorporates an irreversible ingredient and can be associated with the 
interaction of a quantum system with a macroscopic system.

\item The intersection domain ${ K}_p\cap { K}_q$ is empty and similar conditions for large chains.
It corresponds with the case of completely separate systems. This
possibility, as we have discussed before, is also applicable to what
usually we call quantum systems. We can calculate the limit of
non-orthogonality for quantum states. If the maximal Hamilton-Randers
speed is $c_F$, the condition for absence of interferences is given
by the formula
\begin{displaymath}
d_F \rangle  c_F T_{max}.
\end{displaymath}
Because the emission or absorbtion of photons, even the same
notion of photon is from the perspective of deterministic
fisnlerian models, a macroscopical notion, it is desirable to
maintain speed of light as the maximal speed. On the other hand,
because in other way, the introduction of two maximal speeds is
not so desirable. Therefore,
\begin{equation}
d_F \rangle  c T_{max}.
\end{equation}
In the set of compact states, $T_{max}$ is bound by an universal
value $T_0$. This provides the bound $c T_0$ on the Hamilton-Randers
distance in ${ T^* TM}$ for the existence of quantum
interferences for systems defined by compact quantum states. This
also implies a bound for the Riemannian distance in ${ TM}$ and
the distance in ${ M}$.
\end{enumerate}
Our analysis implies that irreversible evolutions is one possible
source for the absence of quantum interferences. 
Other source is the possibility when the system is composed of
strong causal disconnected states. Both mechanisms are independent
and while the first is an attribute of macroscopic objects, the
second one can also be applied at the quantum level.

There are some consequences of the above concepts,

{ Limited range value of the quantum correlations.} That makes at
least theoretically, a notorious difference between our models and
the Quantum Theory: in deterministic finslerian systems there should
exist an universal limit for the quantum interferences r quantum
correlations. The mechanism, in the case of transitions produce by
short chains is clear. Since there is a maximal internal time $T$
and all the speeds are bounded, there is a maximal range of the
quantum correlations. Indeed, this notion defines a causal
structure. This causal structure, however, can be very different
than the usual Lorential structure.

In the case of large chains, one has the possibility that the
quantum state is transported from a quantum state to another quantum
state with different local internal time $T$. In this way a
succession of changes can happens, with slightly different internal
times. The net effect must be therefore a long distance correlation.
However, eventually this must be also finite and the parameter that
fix the range is the maximal Finselrian distance, that is by
definition $L$. Since the speeds are also bounded, this produces a
maximum speed that must be $c$. However, $c$ is also defined as the
fraction $c=\frac{L}{T_0}$ and is universal.

{ Stern-Gerlach type experiments.} We note that the orthogonal
relation is compatible with Stern-Gerlach type experiments because
for having orthogonality it is not necessary that ${ K}_p\cap
{ K}_q=\emptyset$ or generalizations of this condition for large
chains. However, there is an alternative way to reproduce
Stern-Gerlach results without calling to the complex nature of
phases. The associativity rule for orthogonal states capable to
produce Stern-Gerlach type results is
\begin{displaymath}
{ K}_p\cap { K}_r={ K}_{pr}\neq \emptyset,\quad {
K}_r\cap { K}_q={ K}_{rq}\neq \emptyset,\quad { K}_p\cap
{ K}_q\,= \emptyset.
\end{displaymath}
Therefore, using this model a transition of an ontological state can
happens from $p$ to $q$ even if they are orthogonal and their
intersections are empty. The transition from $p$ to $q$ can happen
due to the action of a external field. Therefore complex character
of the transition amplitudes that define the quantum states is not
essential in order to accomplish with Stern-Gerlach
experiments-type. Since we know that complex transitions amplitudes
are physically argued from the analysis of Stern-Gerlach experiments
([20]), we need an alternative description or model for spin. We
propose the following. Two spin states ${ K_p}$ and ${ K_q}$
are orthogonal if ${ K_p}\cap { K_q}=\,\emptyset$. Of course,
there is a basis of orthogonal states $\{{ K}_{(\vec{u},s)}\}$
for each spatial direction $\vec{ u}$. Respect to any other spatial
direction $\vec{z}$ the corresponding orthogonal basis is $\{{
K}_{(\vec{z},s)}\}$. They has to be that ${ K}_{(\vec{u},s)}\cap
{ K}_{(\vec{u},s)}\,\neq \emptyset$. The most natural way to
perform that is to consider a complex projective space, with
elements corresponding to rays in this space. In this space there
are not direct transition between different states of the same
basis. Therefore, if there is such transition it must be through a
secondary or higher order chain. However, even through them, it is
not possible by definition to obtain the corresponding transition.
Note that in the Stern-Gerlach gedanken experiments of [20], there
is a magnetic interaction that change the quantum state as a whole.
The classical field is in this sense {\it persistent}. Now, to allow
the transitions ${ K}_{(\vec{u},s)}\cap {
K}_{(\vec{u},s)}\,\neq \emptyset$ we need the notion second time or
internal time. Through the internal time, the value in the direction
$\vec{u}$ change, while by definition, it keeps constant the value
of $\vec{z}$.

In the case of states $\frac{1}{2}$-spin, a comparison with a
chemical clock is instructive. In a chemical clock, the whole system
change from state B to state R coherently. The transition time is
short compared with the static time. The transition involves
millions of molecules, that due to strong non-linearities changes
simultaneously the state. In a similar way, a given state let say
$|z,+\rangle $ will simultaneously be half of the time in a state $|x,+\rangle $
and half of the time in a state $|x,-\rangle $.

 We should note the following that:

 { One can not
apply Bell's inequalities because the model is applicable to a
single quantum particle}.

{ The Classical Limit.} The classical limit can also be obtained
from deterministic finslerian models in the following way. Let us
assume that for a given point $z\in { K}_p \cap { K}_q$ the
function $d_F(p,z)+d_F(z,q)$ is very large compared with other pair
of distances appearing in the definition of the transition amplitude
(3.1) and compared with $L$. The only possible transitions are for
that the are such that the exponential function is stationary, which
means the variational derivative
\begin{displaymath}
\delta (d_F(p,z)+d_F(z,q))=0.
\end{displaymath}
This is also the condition of being geodesic. Since we are working
with Randers-Berwald spaces, left and right geodesics are the same
because the connection coefficients live in ${ TM}$, although the
metric is not symmetric. Therefore, it is natural to put the
variation
\begin{displaymath}
\delta (d_F(p,z)+d_F(z,q))=\delta (d_F(p,z))+\delta(d_F(z,q))
\end{displaymath}
 with each individual variation of the same order and sign. This is the geodesic condition.
Let us define the classical action $S$ by
\begin{equation}
\frac{d_F(p,z)+d_F(z,q)}{L}\sim \frac{S}{\hbar},
\end{equation}
where $S$ is here the action calculated on the path joining the
extreme points and the distance functions are the length of a path
jointing the points $p, q \in { S^* TM}$. The condition $\delta
(d_F(p,z)+d_F(z,q))=0$ and $L$ very small is therefore equivalent
to the condition that $\delta S=0$ and $\hbar$ very small. This is
the classical limit.
Therefore, classical evolution, defined by the only path that contributes to the integral when 
$\hbar \rightarrow 0$ is equivalent to the 
Hamilton-Randers geodesic evolution in the space ${ TM}$.

\subsection{Rudiments for a Measurement Theory}
Let us consider the quantum state ${ K}_x$ such 
that the point $x\in{ T M}$ is the invariant attractor point. For any other point in 
${ K}_x$, there are local coordinates that will change under the
evolution ${ U}_t$
induced from the geometric evolution $F\to 
h$. We denote these coordinates `` normal" ${\phi}$-coordinates. They could correspond to
what in the literature are denoted by``changeable 
observables". Here they have the meaning of being associated with
non-constant values in the internal evolution of the configuration
of the ontological degrees of freedom.
Coordinates remaining invariant during the $U_t$-evolution (which we call 
co-normal ${\pi}$-coordinates ) are denoted by ``beables" observables, that are well 
defined macroscopically for any particular quantum state. We think
that this distinction that we are doing corresponds to the same
appearing in the literature ([1], [7], [25]).

Now we note the following facts:
\begin{enumerate}
\item  The notion of quantum state represents an objective element
of the physical reality. The quantum state is determined by the
attractor point $x$, but the complete quantum phenomenology is
determined by the sub-set ${ K}_x \subset { TM}$.

\item {Given a point $x\in { TM}$, we can use normal and
co-normal coordinates. The division between normal and co-normal
coordinates is coordinate-free in the sense that any combination
of co-normal coordinates is also co-normal.}

\item {The association of beables with co-normal
coordinates and changeables with normal coordinates depends on the
particular quantum system. There are possible macroscopic
coordinates that could be associated with co-normal coordinates or
with normal coordinates, depending of the quantum state.}
\end{enumerate}
The value of any beable observable is well defined for a given  
quantum state ${ K}_x$ because it is constant during the $U_t$ evolution, while the value of a 
changeable observable is not constant (we denote by beable or changeable these observables). 
We also note that the set of beables is in the general case
non-coincident with the set of ontological coordinates $x$.

The particular value associated with a physical measurement is
defined by a collection of events happening at the Planck scale in
a coherent way. We assume that this is universal, all phenomenon
are determined by events happening at this scale, being by
assumption fundamental. However, it does not mean that all quantum
states have the same size. For instance, the quantum system
describing an electron is very different on size than the one
describing a neutrino. Therefore, in the interaction between
electrons and neutrinos, to which is ultimately reduced any
observation of the neutrino, is very different on size than the
interaction between electron and electron, that is what ultimately
is reduced any observation of an electron state. These events
completely determine the result of macroscopical measurements when
the location in time $t$ is given. The basic dynamical process
corresponding to a measurement involves an interaction which is
both collective (all the degrees of freedom conforming the quantum
system interact with the degrees of freedom of the matter or
photons that conform the measurement device) and complexity (we
have the idea that usual quantum particles are composed by
fundamental degrees of freedom at the Planck scale. The number of
these fundamental degrees of freedom varies but it is usually
large).

What is the process such that the value of a particular coordinate
describing these events at the Planck scale is amplified to be a
macroscopic, observable effect? We can only make the hypothesis
that if the measurement processes are very complex processes that
follow a non-linear dynamics which involves a large number of
basic degrees of freedom, events at the ontological level like
space-time collisions are coordinated to produce macroscopic
collective results that correspond with particular observations.
From this point of view measurement processes are too complicated
to give a reasonable answer in quantitative terms or through an
deterministic and complete detailed evolution process. Therefore
although completely deterministic, a non-deterministic $R$-process
is necessary in the mathematical description of measurement
([22]). However, following our argument, the $R$-process is only
an approximation in our scheme and departures from the usual
instantaneous effects associated with the Quantum $R$-process
should be observed.

We are not able to make numerical prediction, still laking of a
detailed mathematical formalism. However,the mechanism proposed in
these notes differs from the $R$-process. In particular there is
not an instantaneous measurement processes. The processes that we
are considering although essentially very complex, have a finite
speed (finite finslerian speed, that in some sense can exceed the
Lorentzian speed of light but that is in any case finite).
Therefore, we can check our proposal trying to check the apparent
speed on which a measurement phenomena happens. It seems that this
can be achieved observing the apparent instantaneous correlations:
the prediction is that they could be faster than the speed of
light but eventually finite.

A preparation process is associated with a change in the definition of ${ K}_x$. It 
corresponds to a transformation capable to alter the whole quantum state. How this process 
happens? We must agree that a system called ``measurement device" interacts with the quantum 
system. This interaction, happening at the Planck scale, produces a local change in the manifold 
${ S^* T M} $ but in such a way that it changes the global set $({ K}_x)$, changing 
collectively the points defining the quantum state, preparing the
system in other particular quantum state.
 The nature of this global change could be 
associated to the persistence of the interaction between the quantum system and the measurement 
device and universality of the interaction.

A generic combination of beables and changeables $O(\pi, \phi)$ is a changeable 
as well as any combination of changeables only (the exception to this rule can be some special 
combinations as the Casimir operator for spin). In general macroscopic observables are not directly related 
with the $\pi$ or even $\phi $-coordinates. However, due to the property 
of diffeomorphism invariance of the transition amplitudes,
it is possible to use a set of macroscopic observables as normal and 
co-normal coordinates, as soon as the relation between the set of
macroscopic coordinates and
the co-normal and normal coordinates is a diffeomorphism. The existence of 
a splitting in the coordinates of ${ TM}$ is a non-trivial constraint in the possible 
diffeomorphism relating the descriptions at the Planck scale and
usual scales.

\section{Double distance, evolution, time and events}
\subsection{The notion of two-dimensional time}

In this {\it subsection} we address the question of the physical 
interpretation of fundamental notions of the Quantum Theory,
like quantum correlations, entanglement and the 
meaning of the wave function in the contest of deterministic
finslerian models.

Let us start analyzing the interpretation of the quantum state,
or equivalently in our formalism, the interpretation of the 
``transition amplitudes" given by the formula $(3.1)$ between
points of subsets of ${ S^*TM}$.
From the mathematical theory developed in [$3$] appears 
naturally the parameter $t$, running in a compact interval, just marking the evolution of the 
geometry, from Finsler to Riemannian through intermediate geometries with interpolating 
fundamental tensors
\begin{displaymath}
g_{t}=(1-t)g+th,\quad t\in [0,1].
\end{displaymath}
This can be generalized to the expression
\begin{equation}
g_{t}=\frac{1}{T_{max}}((T_{max}-t)g+th),\quad t\in [0,T_{max}].
\end{equation}

This compact time $t$ is different than the external time $s$,
which is non-compact. In addition,
 while the first one is a parameter of the process generating the 
quantum states, the second one is used to describe a macroscopic evolution, classical or quantum 
mechanical. The external time $s$ is independent of the quantum
state. By contrast,
$t$ (because it is compact with maximal value $T_{max}$) is 
related with the nature of the quantum state.
We could assume that it is compact and with 
maximal value $T_{max}$, determined for each particular system as
a intrinsic characteristics of the system. Because the system is
composed by small pair of particles, one moving forward on time
and the other backward on time, the period is bounded by a maximal
period, defined by $L$ and by $c$,
\begin{equation}
T_{max}\langle \frac{L}{c}.
\end{equation}
From the maximal value $T_{max}$, depending on the particular quantum state and a particular 
sub-region of the base manifold ${ T M}$, it follows the locality of the notion of the time 
$t$; being essentially dependent on ${ K}_x$, it could be different for different quantum 
states, that is, different regions of ${ T M}$, although always
under the constraint (4.2).

Equation (4.2) have a significative experimental connotation. Not
only in the field of quantum correlations but also in the field of
the macroscopic quantum interferences. Consider a typical
macroscopic interference experiment with electrons with itself
like the one described in [22](although typically these
experiments are performed with photons). Quantum Mechanics do not
have any limit to the Mach-Zehnder interferences for matter.
However, in our proposal, the existence of finite limits maximal
periods $T_{max}$ implies an eventual limit to these interferences
for matter waves.

In the case of experiments with photons, the maximal finslerian
speed implies also a maximal speed for quantum correlations
between the different paths the photon can take in these
experiments. The limit on these speed is determined by the
finslerian distances in ${ TM}$, the relation with the induced
metric in the space-time and the maximal finslerian speed.

The way the geometry evolves, from Finsler to Riemannian in the
manifold ${ S^* T M}$, is not determined by the relation (4.1).
Indeed, it is possible to use the following relation
\begin{displaymath}
g_{t}=f(s)g+k(s)h,\quad s\in { R},\,\, f+k=1,
\end{displaymath}
with $f,k$ characteristic functions of the system. This argument
proves the need of a dynamical law for the evolution of the
geometry and the practical idea to link the time $t$ with the time
$s$. The dynamical law should be geometrical and the value of the
functions $f,k$ also must have a geometric meaning, linked with
the properties of the quantum state ${ K}_x$.

There are some possible candidates for this dynamics. However,
they should be consistent with the fundamental Poisson structure,
defined by the fundamental hamiltonian ${ H}$.

The Poisson equations,
\begin{equation}
\frac{\partial}{\partial t}g_{ij}= \{g_{ij} ,{ H} \}
\end{equation}
solve the problem of the evolution of the geometry through the
$U_t$-evolution, ones the hamiltonian is specified.

However, If we analyze the number of degrees of freedom we need to
determine the $U_s$ evolution of the system, the specification of
the Hamiltonian is not enough, because contains only $6N$
independent functions, while the complete geometry is given by
$6N(6n+1)/2$ independent degrees of freedom of the metric
${\alpha}$ plus $6N$ of the form $\beta$. One natural candidate is
through a generalization of the Ricci-flow in the Finsler category
(in [23], one possible generalization is investigated). However,
this Ricci flow should be compatible with the intrinsic Poisson
structure given by the equation $(4.3)$. That means an equilibrium
final state for the geometry, given by the average Riemannian
geometry.

It is clear through the arguments presented until now, the existence of two 
different types of dynamics that jointly produce the dynamics of
the quantum systems:
\begin{enumerate}
\item $U_t $-dynamics: every ontological degree of freedom evolves through ${ K}_q$ until 
reaching the equilibrium state $q(s)$. It originates part of the probabilistic character of the 
quantum systems. \item The evolution in the geometry, governed by
the equation (4.3).

\item $U_s $-dynamics: every ontological degree of freedom is replaced by another identical degree 
of freedom in the infinitesimal evolution from $s$ to $s+d s$. The evolution of these $collectives 
$ is defined by the Hamiltonian $(2.5)$.
\end{enumerate}

Comparing with the Quantum Mechanics, the existence of a double
dynamics is a new form of complementing the quantum formalism.
While
usual scales of time assumed of physical measurement processes are so large, $T_{max}$ could 
appear as not detectable because it is usually small for compact
states. In this case,
we can collapse this second making $T_{max}\rightarrow 0$ 
and just say that it corresponds with a macroscopic instant in this limit. Therefore the wave 
function can be written as
\begin{equation}
|\Psi\rangle \, =\int_{{ \Xi}} da \langle a|\psi\rangle |a\rangle
\end{equation}
represents an individual, spread system and has the same interpretation than in the orthodox 
interpretation of Quantum Mechanics.

One exception to this argument is the case of large photon
interferometry. In this case, the time $T_max$ could be large
enough to be detected in the form of deviation from the ideal
quantum correlations.

The line of reasoning presented above could be problematic in case
of non-compact states because $T_{max}$ could be very large.
Therefore we assume, on the basis of the above argument, that all
physical states could be conveniently represented by compact
spaces.

Considering a finite second time $t$, we get a complete, deterministic model as a deeper 
description of the quantum systems.  Reduction of the wave packet is not necessary in the 
formalism when the second time is considered. For example, in a two-slit experiment-type with a quantum 
system, the question for which slit the system pass, the answer we
should give is that for all the possible slits. The key-point to
have a geometric representation (in a generalized phase-space) is
to realize that the notion of ``passing through a slit in some
instant is a macroscopic notion, allowed only when we take the
limit $T_{max}\rightarrow 0$. From the perspective
 of deterministic finslerian models, the relevant question is: at the instant 
$(s,t)$, for which slit is passing the system? The solution
proposed
is that the system pass at this double instant only 
through one of slits. The non-localized character of the quantum
state is due to the fact that the system pass by both slits but at
different double times $(s,t_1)$ and $(s,t_2)$. Quantum systems
are described by a complex system with degrees of freedom at the
Planck scale. However, they are all ``macroscopically localized":
all the degrees of freedom pass through one of the slits.

Form the perspective of deterministic finslerian models, Quantum
Mechanics appears as a remarkable useful tool in dealing with
methods that do not have to treat with these complexes processes,
but with symbolic representations of their macroscopic
descriptions, when the time $T_{max}\rightarrow 0$.
\subsection{Double Distance and Quantum Correlations}

The existence of two distances, the Riemannian and the Hamilton-Randers distance in ${ T M}$ 
could be interpreted physically in the following way. Consider the
metric spaces $({ M},d_F)$ and $({ M
},d_h )$, where the metric distance functions are the induced distances from $({ T M}, d_F )$ 
and $({ TM}, d_h)$ respectively. Let us consider the following 
definition of apparent speeds: for events happening with a difference on time $\Delta s$, there are two 
``apparent macroscopic velocities", $v_F :=\frac {d_F}{\Delta s}$ and $v_R :=\frac {d_R}{\Delta 
s}$ (note that since we are speaking of apparent speeds, we are not allow to use $v_F :=\frac 
{d_F}{\Delta t}$ or $v_R :=\frac {d_R}{\Delta t}$). $v_F$ and $v_R$ could be different, 
but what we know from {\it proposition A.7} is that if one of them is bounded, the other velocity 
should also be bounded.

From the comparison of the Riemannian and Hamilton-Randers volume of the tangent spheres 
([5]), it seems that there is not blow up and speed up of Hamilton-Randers volumes of tangent spheres 
relative to Euclidean volumes. It also seems that this condition implies a relation between the 
distances.

The relation between the quotient of times, implies a possible
more general conformal factor, because the relation
\begin{displaymath}
dt =\frac{\partial t}{\partial s} ds
\end{displaymath}
admits a factor that can be rather large.

Therefore, the apparent quantum correlations appear because we
are using not the correct notion of distance and speed 
between events happening ``inside" the same quantum state ${ K}_x$. The existence of apparent 
speed of order $K\, c $ but not infinity large, is one of the predictions of the theory. Note also 
that this bound is of universal nature, not depending of the internal energy scale or other 
properties of the physical system.

Two technical remarks are in order. Since the distance $d_F$ is
non-symmetric, we need a univalent definition of the distance we
use in the definition of speed. We define the apparent correlation
speed by
\begin{equation}
v_F =min \{\frac{1/2 (d_F(a,c)+d_F(c,b))}{s_{ab}},\,\frac{1/2
(d_F(b,c)+d_F(c,a))}{s_{ab}}\}.
\end{equation}
$c$ is the initial state, producing the entanglement.

We are always calculating distances between points in the space
${ TM}$, using the Finsler structure co-dual of the given dual
Finsler structure $F^*$. This implies, due to the categorization
properties of Randers spaces, an embedding structure in ${ M}$
that is also Randers. We use this induced distance in the
definition of apparent correlations eq. (4.5).

Why we can measure conveniently ``ordinary distances" using the usual Riemannian distance? The 
answer could be given through the introduction of the notion of relative event. This means that 
spatial coordinates and $s$-time $(\vec{x},s)$ can be used to denote two different types of events: events that 
when the difference in the internal time  $t$ between them is small. Then both events could 
happen inside the same quantum state ${ K}_q$. For these events, we should calculate the distance 
with the Finsler measure, as given by the equation (4.5). If the
internal time is large, that means, $t$ is large as $T_{max}$, the use of the 
(pseudo)-Riemannian distance is mandatory because it is the
distance we take when the quantum system reach its equilibrium
state and the metrics are Riemannian.

Following this interpretation, the base space ${ T M}$
appears as an ordering lattice and events are not in ont to one correspondence with what locally can happen. This seems a 
rather breaking fact with the idea to associate Physical Reality with space-time geometry endowed 
with any kind of metric geometry. Indeed, if we should to
implement Quantum Field Theory in the formalism of deterministic
Hamilton-Randers Models, the notion of relative
event presented above and its generalizations could be interesting, because 
different quantum field processes will be associated with different distances between the same points in 
the space-time, using a matrix valued Finsler structure $F_M$.

One consequence of the notion of distance inside of a quantum
state, is the existence of effects which should be slower than
light, when they will propagate theoretically at the speed of
light,
is a consequence of our model. This result comes from the {\it 
equation A.16}: since we have the null integral
\begin{displaymath}
\int _{{ S}_x} \Phi =0
\end{displaymath}
and since $g=h +\Phi$ and $c^2 =g_{ij} v^i v^j$, sometimes the
expected speed will be slower than $c$. The effect can only be
linked with the fundamental Hamilton-Randers character of the
description. It could be suggesting to interpret this variant as
due to the action of the ambient.

The last notion treated in this subsection is convex invariance. It is just the invariance of a 
property by the $U_t$-evolution of the geometry. For example, the
metric $h$ is convex invariance. Any topological property of the
manifold ${ TM}$ is also convex invariance.
Mathematically this notion implies to consider 
the set of dual Finsler metrics over ${ T M}$, ${ T M}_{\mathcal {F}}$, 
${ T M}_{\mathcal {F^*}}$. Given a Riemannian metric $h$, the convex closure $CC(h)\subset { T 
M}_{\mathcal {F}}$ is the maximal subset of all the Finsler functions with average metric $h$. This 
is a convex set. This notion implies to consider the group of transformations of ${ T 
M}_{\mathcal {F}}$ leaving invariant $CC(h)$. Let us call this group quantum symmetry group. The reason 
for this name is that, from the way the quantum state ${ TM}_x$ are defined, they are convex 
invariant. The only change that a quantum symmetry can produce is
a change in the complex phase in the associated vector.
 Therefore, 
the Quantum Symmetry adopts in a natural way, a unimodular group
representation over $\mathcal{H}$.

One possible construction for this unimodular group is the
following:
\begin{displaymath}
\mathcal{U}_{\delta}:\mathcal{H}\to \mathcal{H}
\end{displaymath}
\begin{displaymath}
|x\rangle \, \to e^{\imath 2\pi \frac{d_h (F, F_1)}{diam\,
({ T M}_{\mathcal {F}}})}|x\rangle ,
\end{displaymath}
\begin{equation}
\forall |x\rangle \, \in \mathcal{H}.
\end{equation}
For the definition of this distance and ${diam\, ({ T M}_{\mathcal {F}}})$ we refer to {\it 
Appendix A}. This is defined using a metric structure in $ { T
M}_{\mathcal {F}}$. This metric structure could be useful in the
study of the dynamics of the geometry.

Convex invariance is very useful to understand the relation between Finsler and Riemannian 
geometry and now we show that its inclusion in our scheme makes natural the introduction of 
the complex field ${ C}$ in the axioms of the pre-Hilbert space
associated with the set of quantum states ${ K}_x$.

\subsection{Quantization of Observables}
The quantization procedure of operators that we present consists
of an algebra-morphism between some natural structures that appear
in the mathematical description.

The ontological canonical Poisson structure has the form
\begin{equation}
\{f ,g\}_{(t)}=\sum ^{2N} _{k=1} (\frac{\partial f}{\partial
x^k}\frac{\partial g}{\partial p_k}-\frac{\partial g}{\partial
x^k}\frac{\partial f}{\partial p_k}).
\end{equation}
The internal time $t$ is keep fixed. With this structure, there is
defined an algebra structure in the ring $\mathcal{F}({ T^*
{TM}})$.

The above structure is associated with the $U_t$ evolution and the
ontological degrees of freedom at the Planck scale. However, in
the ordinary description of quantum phenomena it is used a very
different degrees of freedom and a different evolution. Therefore
we have to relate this both descriptions. Since during the $U_t$
evolution, there are coordinates that remain invariant and
coordinates than change, we see that induces the following
decomposition of the poisson structure,
\begin{displaymath}
\{f ,g\}_{(t)}=\sum ^{J} _{l,k=1} (\frac{\partial f}{\partial
\pi^l}\frac{\partial \pi^l}{\partial x^l}\frac{\partial
g}{\partial p_{\pi_l}}\frac{\partial p_{\pi^l}}{\partial
p_{x^k}}-\frac{\partial g}{\partial \pi^l}\frac{\partial
\pi^l}{\partial x^l}\frac{\partial f}{\partial
p_{\pi_l}}\frac{\partial p_{\pi^l}}{\partial p_{x^k}})_t+
\end{displaymath}
\begin{displaymath}
+\sum ^{K} _{k=1} (\frac{\partial f}{\partial
\phi^l}\frac{\partial \phi^l}{\partial x^l}\frac{\partial
g}{\partial p_{\phi_l}}\frac{\partial p_{\phi^l}}{\partial
p_{x^k}}-\frac{\partial g}{\partial \phi^l}\frac{\partial
\phi^l}{\partial x^l}\frac{\partial f}{\partial
p_{\phi_l}}\frac{\partial p_{\phi^l}}{\partial p_{x^k}})_t,\quad
J+K=2N.
\end{displaymath}
where we have used the following relation,
\begin{displaymath}
x^j=x^j(\pi,\,\phi, \,p_{\pi},\,p_{\phi}),\quad
p^j=p^j(p_{\pi},\,p_{\phi},\,\pi,\,\phi).
\end{displaymath}
This coordinate change provides invariance of In this way,
macroscopic variables $\phi$ and the complementary variables does
not commute, as well as the associated moments
\begin{displaymath}
\{\phi,\,\pi\}_{(t)}\neq 0,\quad \{p_{\phi},\,p_{\pi}\}_{(t)}\neq
0.
\end{displaymath}
On the other hand, the fact that all the coordinates ${\pi}$ and
$p_{\pi}$ are well defined at times $s$ justifies the following
condition
\begin{displaymath}
\{\phi,\,\pi\}_{(t)}=0,\quad \{p_{\phi},\,p_{\pi}\}_{(t)}= 0.
\end{displaymath}
Therefore, we should take a reduction of the ontological Poisson
structure to the phenomenological Poisson structure. We propose
\begin{equation}
\{f(\pi,p_{\pi}) ,g(\pi,p_{\pi})\}_{(s)}=\int_{{ K}_{\gamma}}
dz \,e^{\imath \frac{1}{L}(d_F (p,z)+d_F (z,q)-(d_F 
(z,p)-d_F (q,z)))}\,\{f ,g\}_{(t)}.
\end{equation}
From this structure, we have a structure that resembles a to
obtain the macroscopic Poisson structure. The main difference is
the fact that this new structure depends on the particular quantum
state ${ K}$. Therefore, we define quantization using this fact
in the following way. First we introduce the ring of linear
operators on the Hilbert space of quantum states,
$\mathcal{H}:+\{{K}_{\gamma}\}$ and $\mathcal{R}(\mathcal{H})$.
From this homomorphism ring one can define an algebra, considering
the commutator of the operators. This algebra is a Lie algebra and
Jacobi identity comes from associativity property. Then we define
the canonical quantization as the map
\begin{displaymath}
\mathcal{F}( {T^*TM})\to \mathcal{R}(\mathcal{H})
\end{displaymath}
\begin{equation}
\{f,\,g\}_{s}\to \langle K_{\gamma}|[\hat{{
f}},\,\hat{{ g}}]_D\,|K_{\gamma}\rangle ,\quad \forall\,\,
K_{\gamma}\in \mathcal{H}.
\end{equation}
The Dirac bracket is defined as usually
\begin{equation}
[A,B]_D |a\rangle  :=AB|a\rangle  -BA|a\rangle ,\, \,\forall \,\,|a\rangle \in\,\mathcal{H},\,
A,B\in\, Aut(\mathcal{H}).
\end{equation}
The space $\mathcal{H}$ is the Hilbert space described in {\it
subsection} 3.1.

Let us consider an example take one example of how the above
quantization holds. Consider that a pair of the original
ontological variables are also phenomenological variables $f=x^i ,
g=p_j $. Then our quantization relation is just reduced to
\begin{displaymath}
\{x^i,p_j\}=\delta ^i _j\to \int _{{ K}_p\cap {
K}_q}
e^{\imath \frac{1}{L}(d_F (p,z)+d_F (z,q)-(d_F 
(z,p)-d_F (q,z)))}\, \delta ^i _j =
\end{displaymath}
\begin{displaymath}
\delta ^i _j \int _{{ K}_p\cap { K}_q}
e^{\imath \frac{1}{L}(d_F (p,z)+d_F (z,q)-(d_F 
(z,p)-d_F (q,z)))}\, =\delta^i _j ,
\end{displaymath}
because the integrals are normalized to one. The definition (3.14)
is therefore equivalent to
\begin{displaymath}
=\langle p|[{ \hat{x}^i},{ \hat{p}_j}]_D|q\rangle =\delta ^i _j \langle p|q\rangle .
\end{displaymath}
Therefore the Dirac bracket should be $[{ \hat{x}^i },{
\hat{p}_j}]_D =\delta ^i _j { Id}$ that is the canonical
quantization.

We can motivate this quantization in terms of a fundamental,
geometric notions together with statistical considerations through
the following argument. The kernel of the integration could be
simulated as the distribution of a statistical system, but now
assuming an imaginary time. In this way, our quantization could be
completely emergent from a statistical theory at the Planck scale
but assuming an imaginary time through the $U_t $ evolution takes
places. The statistical character of the quantization comes
because quantum states are considered open systems: they are the
result of coordinate structure appearing in  complex systems of
particles, with degrees of freedom scaled associated withal the
Planck scale. However, the quantum state interchanges not only
energy, but also ``matter" with the exterior. Therefore, the
statistical character comes from the treatment of a quantum state
as an open system composed of multitude of particles associated to
the Planck scale.

Secondly, the way we define quantization of operators does not
implies directly an algebra morphism. However, we should prove
that the procedure correctly defines the quantization of
individual classical functions. But at least, it is obvious for
the set of functions that are analytical in the coordinates
$(x,p)$. The relation (3.14) ensures the preservation of the
Jacobi identity and therefore, that the morphism is an algebra
morphism between associative algebras.

Our prescription (3.14) also implies a solution for the ambiguity
in the product operator that appears in canonical quantization. It
defines the quantization through the definition of the expectation
values of operators, that is what really means from a physical
point of view.

For the quantization of operators related with derivations that
are not integrable, let us define the following action on a
sub-manifold ${ K}\subset { S^* TM}$:
\begin{displaymath}
X^i \frac{\partial}{\partial x^i}\to U(X^i)\in
\textrm{Diff(${ S^* K_x}$)},\, U(x^i)=Id-X^i
\frac{\partial}{\partial x^i}.
\end{displaymath}
To the Lie bracket we make correspond the following operator:

\begin{displaymath}
[X^i \frac{\partial}{\partial x^i},Y^i \frac{\partial}{\partial
x^i}]_L f\to \int _{ U^{-1}(Y)U^{-1}(X)U(Y)U(X))({
K}_p\cap { K}_q)}[X^i \frac{\partial}{\partial x^i},Y^i
\frac{\partial}{\partial x^i}]_L (f)
\end{displaymath}
\begin{equation}
e^{\imath \frac{1}{L}(d_F (p,z)+d_F (z,q)-(d_F 
(z,p)-d_F (q,z)))}:=\langle { K}_p|[X,Y]_D f|{ K}_q\rangle f,\,\,\forall
f\in \mathcal{F}_{{ T^* TM}}.
\end{equation}
This implies the required homomorphism between algebraic
structures that we consider as the second type of quantization
(3.13). Typical examples can be operators generating rotations or
other type of transformations in the space.

Finally let us consider the Quantum Hamiltonian from an emergent
point of view. We need to generalize the definition of Hamiltonian
considered in {\it Subsection 2} in order to incorporate the action
on non local states in the sense of being states in ${ TM}$
more general than the fundamental quantum states. The Hamiltonian
element matrix for non-local states is defined by an integration
in a region ${{ K}_p\cap { K}_q}\subset { T^* TM}$,
\begin{equation}
\int _{{ K}_p\cap { K}_q} e^{\imath \frac{1}{L}(d_F (p,z)+d_F (z,q)-(d_F 
(z,p)-d_F (q,z)))}\,\langle { {H}}\rangle \,:=\langle { K}_p|\langle { \hat{H}}\rangle |{
K}_q\rangle .
\end{equation}
In the particular case the states ${ K}_q$ and ${ K}_p$ are
localized states in ${ TM}$, hamiltonian (3.11) is reduced to
hamiltonian (2.5). In the general case, since the regions ${{
K}_p\cap { K}_q}$ are assumed compact, the Hamiltonian is again
bounded from below.

The Hamiltonian operator defined by $(3.16)$ is hermitian, because
the classical Hamiltonian ${ H}$ is real and then the change in
the sing of the exponential function is taken two times, after
conjugation and transposition.

\section{The Quantum S-Matrix}
\subsection{Deterministic Hamilton-Randers Models and S-matrix}
In Quantum Mechanics there is only one dynamics which is linked with experimental data through  
the quantum scattering matrix. The details of the interactions are un-known in this approach to 
the dynamics. In the context of deterministic finslerian
models, two different types of deterministic evolutions are 
present and more detail on the processes is managed, making all
the processes deterministic. However it is also possible to
formulate an unitary matrix that is the quantum mechanical
scattering matrix from the elements appearing in deterministic
finslerian models.

 The ontological scattering matrix
element for a process from the state $a$ to the sate $b$ is
defined by:
\begin{equation}
S_{a b}:=\lim _{s_1 \rightarrow -\infty}\lim _{s_2 \rightarrow
+\infty} \langle a(s_1 )|b(s_2 )\rangle \,.
\end{equation}
Following the usual notions of Scattering Theory, the set of vectors associated with the set of all 
out-states $\{\lim _{s\rightarrow +\infty}\, |b\rangle (s)\}$ conform the pre-Hilbert space or, in 
the case it is complete, the Hilbert space
\begin{displaymath}
\mathcal{H}_{out}:=\{|b(s)\rangle ,\quad s\to 
+\infty \}.
\end{displaymath}
 The 
scattering matrix (5.1) is considered for the case of fundamental,
orthogonal states $|a\rangle $ and $|b\rangle $. 
Analogous considerations for the case of  in-states $\{\lim _{s\rightarrow -\infty} |a\rangle \}$ 
makes natural the introduction of the pre-Hilbert space
\begin{displaymath}
\mathcal{H}_{in}:=\{|a(s)\rangle ,\quad 
s\to -\infty \}.
\end{displaymath}

We show that the above ontological quantum scattering amplitudes generate an unitary quantum 
matrix operator. First, note that $S_{a b}$ is bounded. Then, let
us consider the Fourier transformation of (5.1),
\begin{equation}
S_{\xi _1 \xi _2 }=\lim _{s\rightarrow +\infty}\int _{  M}\int _{ M} d a(s) d b(s)\langle a(s 
)|b(s )\rangle \, e^{\imath a(s)\xi _1}\, e^{\imath b(-s)\xi _2}.
\end{equation}
Developing 
the value $\langle a(s )|b(s )\rangle $ using the geometric Finsler distance, we
obtain
\begin{displaymath}
S_{\xi _1 \xi _2 }=\lim _{s\rightarrow +\infty}\int _{  M}\int _{  M} d a(s) d b(s)\, \int 
_{a_{\gamma}\cap b_{\gamma}} e^{\imath \frac{1}{L}(d_F (a,z)-d_F (z,b)-(d_F (z,a)-d_F 
(b,z)))}\times
\end{displaymath}
\begin{displaymath}
\times e^{\imath a(s)\xi _1}\, e^{\imath b(-s)\xi _2}.
\end{displaymath}
We make the assumption that
\begin{displaymath}
b(-s)=b(s);\quad \xi(-s)=-\xi(s),
\end{displaymath}
recalling the transformation rules for conjugate coordinate and
momentum variables of a point particle.

We promote this matrix with coefficients (5.2) to be
the quantum S-matrix. The measure is 
determined by the phenomenology of the quantum system.

In order to simplify the treatment, let us consider ${ \Xi}
\cong { M}$. This means that physical system have a set of
fundamental quantum states that are labelled by space coordinates.
The orthogonal relations of the exponential function can be
written in the form
\begin{equation}
\int _{  TM} d b(s ) e^{\imath (\xi _l -\xi _m)b(s)} =\delta
(\xi _l -\xi _m),\,\, l,m=1,...,n.
\end{equation}
The inverse relation is written in the form:
\begin{equation}
\int d\xi e^{\imath (a(s)-b(s))\xi}=\delta (a(s)-b(s)).
\end{equation}

The last ingredient used in the proof of unitary property of our
scattering matrix is the decomposition of the unity, that for the
case we are considering is given by the expression
\begin{equation}
I=\int _{{  M}} d\mu (a) |a\rangle \langle a|.
\end{equation}

\subsection{The Quantum S-Matrix is unitary}
 The proof of the unitary relations consist on to perform the
following calculation:
\begin{displaymath}
\int d\xi _{2} S^* _{\xi _2 \xi _1 }S_{\xi _2 \xi _3 }=\int d\xi _{2} \lim _{s\rightarrow 
+\infty}\int _{  M}\int _{  M} d a(s) d b(s)\langle a(s )|b(s )\rangle ^*
\, \times
\end{displaymath}
\begin{displaymath}
e^{ia(s)\xi _1}\, e^{-ib(s)\xi _2}\lim _{s\rightarrow +\infty}\int _{  M}\int _{ M} d c(s) 
d k(s)\langle c(s )|k(s )\rangle \, e^{ic(s)\xi _2}\, e^{-ik(s)\xi _3}.
\end{displaymath}
Re-ordering the $\xi _2$ exponential, performing the integral and using the orthonormal relation 
$(5.5)$ we get
\begin{displaymath}
\int d\xi _{2} S^* _{\xi _2 \xi _1 }S_{\xi _2 \xi _3 }= \lim _{s\rightarrow +\infty}\int _{  
M}\int _{  M} d a(s) d b(s)\langle a(s )|b(s )\rangle ^* \, e^{ia(s)\xi _1}\,
\times
\end{displaymath}
\begin{displaymath}
\lim _{s\rightarrow +\infty}\int _{  M}\int _{  M} d c(s) d k(s)\langle c(s )|k(s )\rangle \, e^{-ik(s)\xi 
_3}\delta (c_{\gamma}-b_{\gamma}).
\end{displaymath}
Integrating the delta function and using hermitian property of the
scalar product, one obtains
\begin{displaymath}
\int d\xi _{2} S^* _{\xi _2 \xi _1 }S_{\xi _2 \xi _3 }= \lim _{s\rightarrow +\infty}\int _{  
M}\int _{  M} d a(s) d b(s)\lim _{s\rightarrow +\infty}\int
_{ M} d k(s)\,\times
\end{displaymath}
\begin{displaymath}
 \langle a(s )|b(s )\rangle  \,\langle b(s )|k(s )\rangle \,e^{-ia(s)\xi _1}\,  e^{-ik(s)\xi _3}.
\end{displaymath}
Using the unitarian condition $(5.5)$
\begin{displaymath}
\int d\xi _{2} S^* _{\xi _2 \xi _1 }S_{\xi _2 \xi _3 }= \lim _{s\rightarrow +\infty}\int _{  M} 
d a(s) \lim _{s\rightarrow +\infty}\int _{  M} d k(s)\,\langle a(s
)|k(s )\rangle \,\times
\end{displaymath}
\begin{displaymath}
 e^{-ia(s)\xi _1}\,  e^{-ik(s)\xi _3}.
\end{displaymath}
From the definition of in-states and taking into account its
orthogonality relation,
\begin{displaymath}
\lim _{s\to \infty }\langle a(s )|k(s )\rangle \,=\delta (a-k),
\end{displaymath}
we get
\begin{displaymath}
\int d\xi _{2} S^* _{\xi _2 \xi _1 }S_{\xi _2 \xi _3 }=\int _{  M} d a(s) \lim _{s\rightarrow 
+\infty}\int _{ M} d k(s)\,\delta (a-k)\,\times
\end{displaymath}
\begin{displaymath}
 e^{-ia(s)\xi _1}\,  e^{-ik(s)\xi _3}=
\end{displaymath}
\begin{displaymath}
= \lim _{s\rightarrow +\infty}\int _{ M} d k(s)\,e^{ik(s)\xi _1}\,  e^{-ik(s)\xi _3}=\delta 
(\xi _1 -\xi _3 ).
\end{displaymath}

An unitary operator can be formulated from the above S-matrix: consider 
the momentum space $\{\xi _a , (,)\}$, where the operation $(,)$
is the scalar product defined in the pre-Hilbert space. Then let
us define
\begin{equation}
({\xi _b}_a , { \hat{S}}\,\xi _b):=S_{\xi _a \xi _b}
\end{equation}
Through this relation it is possible to introduce a link between phenomenology identifying the 
experimental $S$-matrix and $ S _{\xi _a \xi _b }$.

The key point of this proof, that is similar to the standard
derivations ([6]), consists on consider the transitions between
equivalence classes. This is the main idea that we take from the
work of 't Hooft ([7]). Since the set of fundamental quantum
states is considered to be labelled by the space manifold ${
M}$, that implies the integrations are performed in ${ M}$,
except for the decomposition of the unity.

If the set of fundamental quantum states is labelled by a
sub-manifold of ${ M}$, because for instance we consider the
case of quantum states with spin, the domain of integrations
should be performed on a given sub-manifold ${ \Xi}\subset {
M}$. For instance, the definition of the $S$-matrix is:
\begin{displaymath}
S_{\xi _1 \xi _2 }=\lim _{s\rightarrow +\infty}\int _{  \Xi}\int _{  \Xi} d a(s) d b(s)\, \int 
_{a_{\gamma}\cap b_{\gamma}} e^{\imath \frac{1}{L}(d_F (a,z)-d_F (z,b)-(d_F (z,a)-d_F 
(b,z)))}\times
\end{displaymath}
\begin{equation}
\times e^{\imath a(s)\xi _1}\, e^{\imath b(-s)\xi _2}.
\end{equation}
The proof of the unitary property is completely analogous to the
above proof. We use a similar decomposition of the unity in the
space of equivalent classes, and the corresponding orthogonal
relations in the Hilbert spaces describing the ontological states.

\section{Discussion}
\subsection{Generalities}
There are several ideas that have been introduce in this work and
that deserve some additional words. The first one
is the existence of information loss and dissipative geometric dynamics. 
In our approach, the dissipative dynamics is encoded 
in the evolution of the metric structure of the phase space from
Finsler to Riemann in the space ${ TM}$. This evolution
contains the notion of internal time as a parameter driving the
evolution of the geometry, which is different than macroscopic
time. Since the geometric evolution is dissipative (there are
infinite Finsler structures evolving to the same Riemannian
structure) this implies is an intrinsic arrow for the internal
time.

Therefore irreversibility is incorporated at the fundamental level
in our theory. However because this reversibility affects the
evolution in the internal time, the laws of Physics, describing
evolution on the external time or parameter, are not affected.
Indeed, one can formulate laws where no internal parameter appears
and the same kind of fundamental irreversibility associated with
the internal evolution of the geometry could appear.

Associated with the external and internal time, there is the
notion of double distance. Because there is a Riemannian structure
coming from averaging a Hamilton-Randers structure, there are associated
different notions of distances between points: points in the space
can potentially be associated with two types of events. First,
events associated with degrees of freedom at the Planck space. In
this case, the distance between them is associated with a
Hamilton-Randers distance. If the points are associated with quantum
measurements of a system, the distance is the space-time distance,
which is a Riemannian distance associated with a Lorentz structure
on the space-time ${ TM}$. This pseudo-Riemannian structure
must be constructed from fundamental Riemannian structure that
appears as a average of the Hamilton-Randers structure on { TM}.
Several possibilities are discussed later.

The same mechanism 
produces a split of the null ``equilibrium hamiltonian", appearing
a positive part, that corresponds 
to matter (including graviton) and a negative part,
which must be associated with the 
gravitational energy ([2]). Since the exact Hamiltonian can be
decomposed as
\begin{displaymath}
{ H}=\langle { H}\rangle +\delta { H}.
\end{displaymath}
Because the operator $\langle { H}\rangle $ is bounded from below, it must be
defined positive. Therefore, the Operator $\delta { H}$ is
always positive defined. Therefore the above splitting is natural,
but the identification as before seems much more arbitrary. It
seems that there is not any other candidates.

Although the ontological dynamics that we have introduced happens 
at the Planck scale, some testable consequences are proposed.
However, for the moment all of them have a qualitative character.
Improved quantum correlation 
experiments can test the speed of the quantum correlations. From
our theory one concludes the  existence of bounds for this speed.
Although faster than the speed of light in vacuum, because the
quantum correlations are associated with events at the Planck
scale that are separated in the internal-time and since the
(Hamilton-Randers) speed is bounded, the observable correlation speed
must also be bounded.

Other effect follows from the general theory developed: the 
apparent delay of particles propagating theoretically with the
speed of light in vacuum $c$. It is a consequence of the geometric
mechanism generating the dissipative dynamics. This fact is a
special feature of our scheme and differentiates it from others
attempts to obtain a basic dynamics at the Planck scale.
\begin{center}
***
\end{center}

We can compare the prediction of the maximal acceleration of ref.
[8] with the work of Caianiello et Al.  
on maximal acceleration, reported in ref. [9]. If the origin of
maximal acceleration is a fundamental dynamics at the Planck scale, it is 
rather difficult to check the maximal acceleration because it will
be very large: if the mass scale is the Planck scale, then maximal
acceleration has the universal value:
\begin{equation}
A_{max}\sim 10^{52} m/s^2.
\end{equation}
But if the mass $m$ is of the order of energy scale of the 
physical system that is accelerated, then the maximal acceleration is given by a similar formula, 
Caianiello's formula:
\begin{equation}
A_{max}=\frac{2mc^3}{\hbar}.
\end{equation}
In this case experimental tests could be possible as is discussed
in the literature.

However, we do not find any reason in our scheme to 
link the maximal acceleration with the scale of the system; from
the point of view of Hamilton-Randers deterministic systems a universal
and fundamental maximal acceleration is more natural than a
dependence on the mass. However, we should introduce the concrete
value of the fundamental energy-mass scale. This scale is
associated with the vacuum structure at the Planck scale,
consisting on a distribution of minimal density for pairs of point
particles at one point having minimal mass. If the vacuum
structure provides a minimal mass (and not the Planck scale as
energy scale), this also provides a universal maximal
acceleration, that is relative small compared with (6.1). The
minimal mass presently know for matter is the neutrino mass, and
therefore, from this perspective, the maximal universal
acceleration could be
\begin{equation}
A_{max}\sim \frac{2m_{\nu} c^3}{\hbar},
\end{equation}

Comparing Caianiello's Quantum Maximal 
Acceleration (6.2) with our formula (6.3), should provide an
indirect
 check of Quantum Mechanics against deterministic 
finslerian models; Caianiello's maximal acceleration, depending on
the mass of the system,
could be so different from ours 
universal maximal acceleration (6.3), that this could also be a
test of our theory.
\begin{center}
***
\end{center}

The above argument can also be extended to the problem of the
cosmological constant and the associated coincidence problems. If
the vacuum is formed by pair of associated particles (not-really
punctual, but with some extension ([10]) in order to accomplish
with the ergodic hypothesis in a finite time in the sub-manifold
of ${ S^* TM}$ subject the Legendre transformations), the mass
of the pair of particles should be distributed, defining an
averaged density. Let is also take the relation of 't Hooft
relating the periods of the limit cycles with the energy ([7]),
\begin{equation}
E=\frac{\hbar}{T_{max}}.
\end{equation}
If the particles are neutrinos, we obtain a vacuum density energy
([11])
\begin{equation}
\rho_o =\frac{2(m_{\nu} c^2)^4}{4\pi(\hbar c)^3}.
\end{equation}
These formula can provide a possible solution for the cosmological
constant problem and the coincidence problems. We will consider
this topic more extensively in ref. [11].

\subsection{Space-Time Phenomenological Geometries}
One of the problems that we get is to relate the formalism of the
mechanical models describing deterministic systems with the
structure of the space-time, as is formulated and used in Field
Theory and General Relativity. This problems have sense, because
one can consider that our defining universe consists of a main
system and  surrounding mater. Then the surrounding mater can be
describe by an effective stress-energy momentum tensor. The system
under consideration is defined (without considering internal spin
or other degrees of freedom) through a $6$-dimensional tangent
space manifold ${ TM}_3$ and the $2$-dimensional time manifold
${ R}\times { S}^1$.

We assume for the Euclidean space-time a product structure of the
type ${ M}_5={ M}_3\times { C}$, where ${ C}$ is an
undefined cylinder. Because we assume that ${ M}_3$ is
complete, the product is complete as well. We can define, using
the averaged, a positive defined metric on ${ M}_3$. If
$\tilde{g}$ is the horizontal lift (also known as Sasaki-type) of
the fundamental tensor $g$ associated with the Finsler structure
and $\iota_u(X)$ is the horizontal lift of a vector field $X$,
defined using the non-linear connection of the Berwald type
associated to the Randers structure. Then, the Riemannian metric
on ${ TM}_3$ is formally defined as
\begin{displaymath}
h(X,Y)=\int_{{ T}_x{ M}}
\tilde{g}(\iota_u{X},\iota_u{Y})du,\quad u\in { T}_x{ TM},
\quad x\in{ TM}.
\end{displaymath}

The Randers structure can be such that both the associated
Riemannian metric $\alpha$ and the $1$-form $\beta$ depends on the
external time $t$. Therefore, the metric on ${ M}_3$ is given a
tensor $h(x,t)$ that can be read from the fundamental tensor of a
Randers structure,
\begin{displaymath}
g_{ij}(x,y,t)=\Big(\frac{F(x,y,t)}{\alpha(x,y,
t)}a_{ij}(x,t)+l_il_j\Big)-\frac{F(x,y,t)}{\alpha(x,y,t)}\tilde{l}_i\tilde{l}_j,\quad
l_i=\frac{\partial F}{\partial y^i},\quad
\tilde{l}_i=\frac{a_{ij}}{\alpha}{y^j}.
\end{displaymath}
Using standard formulae of Randers spaces and if the averaged is
performed on the manifold defined as $\alpha=1$, one can show that
the averaged metric depends on the moments of $f$ on $y$ up to
third order,
\begin{displaymath}
h_{ij}(x,t)=a_{ij}+a_{il}a_{jk}\,\langle y^ly^k\rangle  + \beta_i\beta_j
+a_{il}\beta_j \langle y^l\rangle +a_{jl}\beta_i \langle y^l\rangle \,+a_{ij}\beta_k\,\langle y^k\rangle -
\end{displaymath}
\begin{equation}
-a_{ik}a_{jl}\beta_m\,\langle y^m\,y^j\,y^k\rangle .
\end{equation}
This provides us a with a natural metric still in a
$6$-dimensional manifold ${ TM}$. In order to have a Riemannian
metric on the manifold ${ M}$ we consider the isometric
embedding
\begin{displaymath}
\zeta:{ M}\to { TM}
\end{displaymath}
\begin{displaymath}
\zeta^*(h)(X,Y)=h(\zeta_*(X),\zeta_*(Y))
\end{displaymath}
using the above metric $h$. This construction provides a natural
metric which is on the $3$-manifold ${ M}$ and that we still
call $h$. At

Now we consider the Riemannian metric structure for the
$2$-dimensional time manifold. Since it is a cylinder and it has a
natural metric (the embedded from ${ R}^3$ we can consider the
corresponding metric. In polar coordinates, the metric of the
cylinder is just
\begin{displaymath}
d{ s}^2=\rho^2(x,t)\,ds^2+dt^2.
\end{displaymath}
Since this metric does not depend on $\theta=s$, the averaged on
$s$ coincides with the original metric. This averaged is on the
fiber ${ S}^1\subset { S}^1\times { R}$. Furthermore, we
can re-scale the metric, in such way that the integral along
$theta$ does not depend on the external time:
\begin{equation}
d{ s}^2=ds^2+\,\frac{1}{\rho^2(x,t)}dt^2.
\end{equation}
Therefore, consider the sub-manifold { R} of the cylinder {
C} and the metric induced. The corresponding block of the
Euclidean space-time metric has the form
\begin{displaymath}
\frac{1}{\rho^2(x,t)}dt^2.
\end{displaymath}
Therefore, the Euclidean space-time metric is given by the
following expression
\begin{displaymath}
ds^2_4=\frac{1}{\rho^2(x,t)}dt^2+\big(
a_{ij}+a_{il}a_{jk}\,\langle y^ly^k\rangle  + \beta_i\beta_j +a_{il}\beta_j
\langle y^l\rangle +a_{jl}\beta_i \langle y^l\rangle \,+a_{ij}\beta_k\,\langle y^k\rangle -
\end{displaymath}
\begin{equation}
-a_{ik}a_{jl}\beta_m\,\langle y^m\,y^j\,y^k\rangle \big)dx^i\otimes dx^j,\quad
i,j=1,2,3.
\end{equation}
In order to define from the Euclidean space-time metric the
Lorentztian space-time metric we need a time-like global vector
field. Then, if the vector field is on the time direction $t$, the
Lorentz metric is
\begin{displaymath}
ds^2_4=-\frac{1}{\rho^2(x,t)}dt^2+\big(
a_{ij}+a_{il}a_{jk}\,\langle y^ly^k\rangle  + \beta_i\beta_j +a_{il}\beta_j
\langle y^l\rangle +a_{jl}\beta_i \langle y^l\rangle \,+a_{ij}\beta_k\,\langle y^k\rangle -
\end{displaymath}
\begin{equation}
-a_{ik}a_{jl}\beta_m\,\langle y^m\,y^j\,y^k\rangle \big)dx^i\otimes dx^j.
\end{equation}
Note that due to the embedding This equation is a definition of
space-time metric. The way in which we have defined the Euclidean
space-time metric implies that $h(e_0,e_i)=0,\, i=1,2,3$. This is
also true for the Lorentz metric (6.2.4). However, this last
formula relies on the choice of the time-like vector field; In
general, it is not the case and the selection of the time-like
vector field is such that the products $h(e_0,e_i)\neq 0$,
providing a Lorentz metric.

This Lorentz metric (6.2.4) depends on the following parameters:
$a_ij$, which is equivalent to $21$ independent parameters, a
vector field $\beta$, which provides $6$ independent parameters
and a vector field defined in the space-time which is equivalent
to $3$ parameters. However, due to simplicity, it is better to
consider that $a_{ij}$ is flat (zero parameters) and that the
global vector field is fixed by the global defined $1-form$
$\beta_i$, $i=4,5,6$. The fourth component of the time-like vector
field is then given by $W^0=\sqrt{m^2-\sum a^{ij}\beta_i
\beta_j}$, where $a^{ij}$ is the inverse matrix of $a_{ij}$.
Therefore, one has to be sure that the vector field $W$
constructed in this way, is still time-like. This reduction in the
number of parameters means that the metric $h$ has the possibility
to be determined phenomenologically using a space-time field
theory invariant under diffeomorphisms.

The recovering of a space-time metric defined by a diffeomorphic
invariant is under the requirement that $a_{ij}$ is flat, while
$\beta$ is dynamical. The general situation where the metric
$a_{ij}$ is dynamical eventually produces a generalization of the
model of the phenomenological geometry. Then, more conditions are
required in order to determine completely the dynamics of the
complete system.

\appendix

\section{Mathematical Appendix}

\subsection{Pseudo-Finsler Geometry.}

A pseudo-Finsler structure or non-degenerate Finsler or
semi-Finsler structure is the following:
\begin{definicion} (Beem's definition)
A non-degenerate Finsler structure $F$ defined on the
$n$-dimensional manifold ${ M}$ is a real function  $F:{
TM}\rightarrow { R}$ such that it is homogeneous of degree $2$
in $y$, smooth in the split tangent bundle ${ N}={
TM}\setminus \{0\}$ and the hessian matrix
 \begin{equation}
g_{\mu \nu}(x,y): =\frac{1}{2}\frac{{\partial}^2 F
(x,y)}{{\partial}y^{\mu} {\partial}y^{\nu} }
\end{equation}
is non-degenerate in ${ N}$, being ${ N}={
TM}\setminus{0}$. The particular case when the manifold is
$4$-dimensional and $g_{\mu \nu}$ has signature $(+,+,+,-)$ is
called Finsler space-time.
\end{definicion}
$g_{{\mu}{\nu}}(x,y)$ is the matrix of the fundamental tensor
$g=g_{\mu\nu}dx^{\mu}\otimes dx^{\nu} $. A generalized Finsler
structure drops from this {\it definition A.1.1} the homogeneity
requirement.

Pseudo-Riemannian structures are the most important examples of
pseudo-Finsler structures. Another important subcategory are
pseudo-Randers structures,
\begin{ejemplo}
A Randers space is characterized by a Finsler function of the
form:
\begin{equation}
F(x,y)=\alpha(x,y)+\beta(x,y),
\end{equation}
where $\alpha (x,y):=a_{ij}(x)y^i y^j$ is a Riemannian metric and $\beta(x,y):=\beta _i (x)y^i$. 
The requirement of being $g_{ij}$ positive definite implies the 1-form $(\beta _1,...,\beta _n )$ 
is bounded with the Riemannian metric $a_{i j}$:
\begin{displaymath}
\beta _i \beta _j a^{ij}\leq 1.
\end{displaymath}
\end{ejemplo}
\begin{definicion}
Let $({ M}, F)$ be a non-degenerate Finsler structure and
$(x,y,{ U})$ a local coordinate system on ${ TM}$. The
Cartan tensor components are defined by the set of functions,
\begin{equation}
{ A}_{\mu \nu \rho }=\frac{1}{2}
\frac{\partial g_{\mu \nu }}{\partial y^{\rho}},\quad 
\mu,\nu, \rho =0,...,3.
\end{equation}
\end{definicion}
The components of the Cartan tensor are zero if and only
if the Finsler space-time $({ M},F)$ is a 
Lorentzian structure. In the general case, our definition of the
Cartan tensor is an homogeneous tensor of degree $-1$ in $y$. The
Cartan tensor is defined to be
\begin{equation}
A(x,y):=\frac{1}{2} \frac{\partial g_{ij}}{\partial y^{k}}\,
\frac{{\delta}y^i}{F} \otimes dx^j \otimes dx^k=A_{ijk}\,
\frac{{\delta}y^i}{F} \otimes dx^j \otimes dx^k .
\end{equation}

By pseudo-Finsler geodesics we mean the parameterized curves in
${ M}$ that are extremal of the pseudo-Finsler functional arc
length. They are solutions of the differential equations (in the
case of unit parameterized pseudo-Finsler geodesics)
\begin{equation}
\frac{d^2  x ^i}{ds^2} +\gamma^i _{jk} (x,y)\frac{dx^k}{ds}
\frac{dx^j}{ds}=0,\,\, i,j,k=1,...,n.
\end{equation}
The connection coefficients of the non-linear Berwald connection
are defined by the derivative of the spray coefficients:
\begin{equation}
\Gamma^i _{lm} =\frac{\partial^2 }{\partial y^l \partial y^m
}(\gamma^i _{jk} (x,y)y^k y^j).
\end{equation}
These connection coefficients define a non-linear connection in
the bundle ${ TN}$. The semi-spray coefficients are defined by
\begin{equation}
G^i =\gamma^i _{jk} (x,y)y^j y^k.
\end{equation}

Using the coefficients ({ D.6}), one obtains a splitting of
{ TN} in the following way. Let us consider
 the local coordinate system $(x,y) $ of the manifold ${ TN}$ and
 an open subset ${ U}\in { N}$. Then, a tangent basis for
 ${ T}_u { 
N}$ is given by the local distribution
\begin{equation}
\{ \frac{{\delta}}{{\delta} x^{1}}|_u ,...,\frac{{\delta}}{{\delta} x^{n}} |_u, 
\frac{\partial}{\partial y^{1}} |_u,...,\frac{\partial}{\partial
y^{n}} |_u\},\quad
 \frac{{\delta}}{{\delta} x^{\nu}}|_u =\frac{\partial}{\partial
x^{\nu}}|_u -N^{\mu}_{\nu}\frac{\partial}{\partial y^{\mu}}|_u .
\end{equation}
The coefficients $N^{j}_k$ are defined in function of the
coefficients $G^i$ by
\begin{displaymath}
G^i=N^i _j y^j,\quad i,j=1,...,n.
\end{displaymath}
The set of local sections $\{ \frac{{\delta}}{{\delta} x^{1}}|_u 
,...,\frac{{\delta}}{{\delta} x^{n}}|_u,\, u\in { U} \} $
generates the local horizontal distribution $\mathcal{H}_U $,
while $\{ \frac{\partial}{\partial y^{1}}|_u ,..., 
\frac{\partial}{\partial y^{n}}|_u, u\in { U} \}$ the local
vertical
distribution 
$\mathcal{V}_U$. The subspaces ${ \mathcal{V}}_u $ and ${
\mathcal{H}}_u$ are such
that the splitting of ${ 
T}_u { N}$ holds:
\begin{displaymath}
{ T}_u { N}=\mathcal{V}_u \oplus \mathcal{H}_u ,\, \forall
\,\, u\in { U}.
\end{displaymath}
This decomposition is invariant by the right action of ${
GL}(n,{ R})$ and defines a connection in the reduced bundle
${ E}({ TN},{ Gl}(n,R))$ of the tangent bundle ${
TN}\rightarrow { N}$, which is a connection in the sense of
Ehresmann in the principal bundle ${ TN}({ N}, {
GL}(n,{ R}))$.

An example of a pseudo-Finsler connection is the Berwald
connection. It is a linear connection defined on the vector bundle
$\pi^*{ TM}\to { M}$,
\begin{proposicion}
Let $({ M}, F)$ be a pseudo-Finsler structure. Then the
$g$-compatibility condition of the Berwald connection is:
\begin{equation}
{\nabla}^b_{V(X)}\pi^* g=2A(X,\cdot,\cdot),
\end{equation}
\begin{equation}
{\nabla}^b_{H(X)} \pi^* g= -2\nabla^b _ {l}A(\cdot,\cdot,X),\quad
l=\frac{y^i}{F}\frac{\partial }{\partial x^i} .
\end{equation}
\end{proposicion}
\begin{proposicion}
Let $({ M}, F)$ be a pseudo-Finsler structure. Then the linear
Berwald connection is torsion free:
\begin{enumerate}
\item Null vertical covariant derivative of sections of ${\pi}^*
({ TM})$: let $\tilde {X} \in { T}_u{ N}$ and $Y\in {
\pi^*}({ \Gamma M})$, then
\begin{equation}
{\nabla}^b_{{V(\tilde{X})}} {\pi}^* Y=0.
\end{equation}
\item  Let us consider $X,Y\in { TM}$ and the associated vector
fields with horizontal lifts in { TN} with components $X^i $
and $Y^i $, $\iota({X})$ and $\iota({Y})$ respectively. Then
\begin{equation}
{\nabla}^b _{\iota({X}}) {\pi}^* Y-{\nabla}^b_{\iota({Y})}{\pi}^*
X-{\pi}^* ([X,Y])=0.
\end{equation}
\end{enumerate}
\end{proposicion}
Given a non-linear connection on a vector bundle ${ TTM}$, it
is always possible to define lifts. In particular, as we defined
before, horizontal lifts of tangent vectors. But one can also
define lifts of other structures, like fundamental tensors. In
particular, one can define the horizontal lift of a Riemannian
structure $h$, which is the Sasaki-type metric
\begin{displaymath}
\left( \begin{array}{cc}
h(x) & 0\\
0 & h(x)
\end{array}\right).
\end{displaymath}

\subsection{Averaged Structure}
The simplest definition of averaged of a function of the
coefficients that we are using is
\begin{equation}
 \langle f\rangle  :=\int _{{ S}_x} |\psi (x,y)|^2 f ,
\end{equation}
$|\psi (x,y)|^2$ is the weight function on the sphere ${ S}_x$.

In the case of smooth Finsler structures the coefficients $\{ h_{ij},\, i,j=1,..,n \} $
 are smooth functions over ${ M}$. They are the components of a Riemannian metric in ${ M}$.
Therefore one has the following
\begin{proposicion}
Let $({ M},F)$ be a Finsler structure. Then the functions
\begin{equation}
h_{ij}(x):=\langle  g_{ij}(x,y)\rangle ,\quad  \forall \,\, x\in { M}
\end{equation}
are the components of a Riemannian metric on ${ M}$ such that in a local basis $(x,{ U})$ 
the metric can be written as
\begin{equation}
h(x)=h_{ij} dx^i\otimes dx^j.
\end{equation}
\end{proposicion}
Associated with the homotopy parameter $t$, there is a whole set
of Hamilton-Randers structures, defined by a fundamental tensor $g_t$,
\begin{displaymath}
g_t= th\, +(1-t)g.
\end{displaymath}

Given a norm $\|,\|$ on the tangent space ${ T}_x { M}$ the distance between 
two points is given by the expression
\begin{displaymath}
d(p,q )=inf\, \{\int ^{q }_{p } \|T(\gamma,\dot{\gamma})\| \},
\end{displaymath}
where $\gamma$ is a measurable path joining the points $p$ and
$q$.
The right-center of mass of a compact sub-set $K\subset { M}$ is defined as the 
point minimizing the function:
\begin{displaymath}
CM_r : K\to { R}
\end{displaymath}
\begin{displaymath}
p\to \int _{K} d^2 _F (p,a)\, da.
\end{displaymath}
$da$ is a measure defined on $K$. A similar notion can be
considered for the integrand $d^2 _F (a,p)$. The corresponding
point minimizing the integral is a functional defined by $CM_l$
(the left center of mass function).

The same construction can be done for the interpolating metric $g_t$. Therefore, let us 
consider the symmetric function:
\begin{displaymath}
\frac{1}{2}(C M_r +C M_l)(t):K\to { R}
\end{displaymath}
\begin{equation}
p\to \frac{1}{2}(\int _{K} d^2 _t (p,a)\, da +\int
_{K} d^2 _t (a,p)\, da).
\end{equation}
From the definition of the interpolating metric $g_t$, the above integral can be decomposed in a 
Riemannian and non-Riemannian components, denoted by $CM_1$ and
$\delta CM$:
\begin{displaymath}
\frac{1}{2}(C M_r +C M_l)(t)=C M_1 +\delta C M,\quad C M_1
(t):=t\int _{K} d^2 _h (p,a)\, da ,
\end{displaymath}
\begin{equation}
\delta C M:=\frac{1}{2}(1-t)\big(\int _{K} d^2 _t (p,a)\, da +\int
_{K} d^2 _t (p,a)\, da \big).
\end{equation}
The center of mass is a zero of a vector field ([12]).
Because the zeroes of smooth vector fields are invariant under continuous transformations, it 
follows that the differential condition $\frac{\partial}{\partial x^i} C M_1 =0$ implies also
the differential condition $\frac{\partial}{\partial x^i} (\frac{1}{2}(C M_r +C M_l)(t))=0$ has a solution,
although they are not necessarily the same. By a 
theorem of E. Cartan, there is a point such that $\frac{\partial}{\partial x^i} C M_1 =0$. Therefore 
we proved the following
\begin{teorema}(Existence of the center of mass)
Let $({ M},F)$ be a Finsler manifold and let 
$K\subset { M}$ be a compact sub-set. Then there is a point
$p_1 $ minimizing the function
\begin{displaymath}
\frac{1}{2}(CM_r +CM_l )(t): K\to { R}
\end{displaymath}
\begin{displaymath}
p\to \frac{1}{2}(\int _{K} d^2 _t (p,a)\, da +\int
_{K} d^2 _t (a,p)\, da)
\end{displaymath}
\end{teorema}
Analogous results hold for the functions $CM_l$ and $CM_r$. We
remark that the existence of these zeroes has been used in the
construction of the pre-quantum states.

The next result is relevant for the definition of relativity group
that is compatible with a underlying Finsler structure,
\begin{proposicion}
Let $({ M},F)$ be a Randers structure and $({ M}, h)$ the associated Riemannian structure. 
Then the isometry group of $F$ is a sub-group of the isometry group of $h$, $Iso(g)\subset 
Iso(h)$.
\end{proposicion}
{ Proof:} From the formula for the metric $h$ it is clear that any linear transformation 
leaving invariant $F$ or $g$ should also leave $h$ invariant, because it is given in terms of $F$ 
and $g$, including the integration domain.\hfill$\Box$

The interpretation of this statement is the generation of symmetry
when we consider the averaged dynamics. Therefore, if the dynamics
at the Planck scale is defined by a function with less symmetry
that standard physical systems (which contains for instance the
rotational symmetry). Testable Bell's inequalities are based on
rotational symmetry (the states are represented as elements of
vector representations of vector fields), there is not reason to
expect Bell's inequalities hold at the Planck scale if the
rotational symmetry is generated through the averaged.

The following proposition shows that the Finsler and Riemannian
distances don't blow-up between each other,
\begin{proposicion}
Consider the average of the metric coefficients $\langle g_{ij}\rangle $ and the line integral $\int ^q _p 
(g_{ij}T^i T^j)^{\frac{1}{2}}$ along a path joining the points $p$
and $q$. Then,
\begin{displaymath}
\int ^q _p \,dt(\langle g_{ij}(x(t),u(t))\rangle _{u(t)}  T^i T^j 
)^{\frac{1}{2}}\,  \simeq \, \int ^q _p (g_{ij}(x(t),u(t))T^i T^j
)^{\frac{1}{2}}\,\,\,\forall u(t)\in { T}_{x(t)}{
M},\,\,t\in [0,1].
\end{displaymath}
\end{proposicion}
The meaning of the above equivalence relations is that both expressions are similar: if 
one of the distances is bounded, then the other is bounded as
well.

The next result provides an example of comparison  between the
Finsler and the Riemannian distance, following the above {\it
proposition}
\begin{proposicion}
Let $({ M},F)$ be a Finsler structure. If the fundamental tensor $g $ is decomposed as 
$g=h+\Phi$ and $\Phi$ is bounded by $g$ and $-g$, then
\begin{equation}
2g\rangle h .
\end{equation}
\end{proposicion}
{ Proof:} we can write the fundamental tensor in the form
\begin{equation}
g_{ij} y^i y^j = h_{ij }(x)y^i y^j\, + \Phi _{ij} (x,y) y^i y^j
\end{equation}
and using the average operation,
\begin{displaymath}
\int _{{ S}_x} g\,=\int _{{ S}_x} h +\int _{{ S}_x} \Phi
=h
\end{displaymath}
and therefore,
\begin{equation}
\int _{{ S}_x} \Phi =0.
\end{equation}
This implies the existence of negative contributions to the
integral of terms containing $\Phi$. Then
\begin{displaymath}
g_{ij} y^i y^j - h_{ij }y^i y^j\, =\Phi _{ij} y^i
y^j\Longrightarrow 2g\rangle h.
\end{displaymath}
This gives a strong bound for $g$ in terms of $h$, when the
differences are small.\hfill$\Box$

We have introduced the notion of convex invariance in ref. [3].
\begin{definicion}
Let $({ M},F)$ be a Finsler structure and consider the 1-parameter family of Finsler structures 
with fundamental tensors $g_t =(1-t)\, g +t\,\langle g\rangle $. A property will be called convex invariant if it 
holds for every $t\in [0,1]$.
\end{definicion}
Associated with $t$ we have not only a Finsler metric $g_t$ but also other geometric objects like 
connection and the associated curvatures, that comes from convex
combinations of the initial connection and the averaged
connection, similarly to the interpolating fundamental tensors. We
call these objects Hamilton-Randers.
\begin{definicion}
Consider an arbitrary Finsler structure and the associated averaged Riemannian structure $({ M},h)$.
A property will be called Riemannian if 
it is completely specified from the Riemannian structure $({
M},h)$. A property is Hamilton-Randers if it is determined by the
Finsler structure.
\end{definicion}
An example of convex invariant property is a topological property, not depending of the metric, 
but only on the underlying topology of the manifold ${ M}$.

The basic result used to translate results from Finsler geometry to Riemannian geometry is the 
following theorem:
\begin{teorema}
Let $({ M},F)$ be a Finsler structure. Then a Riemannian property is convex invariant iff it is 
a Finsler property.
\end{teorema}
This property implies an invariance under a generalized $U_t$-dynamics
 of the Riemannian properties. Since the $U_t$-dynamics is fundamental
 in our construction of the quantum states. We also remark that the 
notion of convex invariance is of fundamental importance in the treatment of Finsler and 
Riemannian geometries as different aspect of a common ``geometry"
([3]).

It is clear that the above property justifies the study of the space ${  M}_{\mathcal {F}}$ 
of the Finsler structures over ${  M }$. Therefore, the introduction of a distance function in the 
manifold ${ M}_{\mathcal{F}}$ becomes interesting.
In particular, we adopt here the construction of Ref. ([12]). First 
note that to a given Finsler structure $({  M}, F)$ it is always possible to associate a 
Sasaki-type riemannian structure $({ T M}, g\oplus g)$. This association
 suggests a smooth embedding of 
${  M}_{\mathcal {F}}$ in the set of Riemannian structures
${ (T  M)}_{\mathcal {R}}$,
\begin{displaymath}
{  M}_{\mathcal {F}}\to { (T M )}_{\mathcal {R}}
\end{displaymath}
\begin{displaymath}
F\to g\oplus g.
\end{displaymath}
The construction of Michor is applicable to this associated Sasaki-type metric, implying the 
following definition for the Riemannian metric $G_{\tilde{g}}$,
\begin{equation}
G_{\tilde {g}}(F_1 ,F_ 2)=\int _{{ TM }} dvol(\tilde{g})\, Tr (\tilde{g}^{-1}g_1 
\tilde{g}^{-1}g_2).
\end{equation}
This is a direct adaptation of the construction found in [$12$]. We should remark that ${ M }$ 
is not necessarily compact. This metric is invariant under diffeomorphism, symmetric and positive 
definite.

Finally, the notion of diameter in ${ K}\subset { M
}_{\mathcal{F}}$ is given by
\begin{equation}
diam ({ K})=inf\{d_{\tilde{g}}(F_1 , F_2),\,\, F_1 , F_2 \in
{ K }\},
\end{equation}
where the metric distance $d_{\tilde{g}}(F_1 , F_2)$ is associated with the metric 
$G_{\tilde{g}}(F_1 , F_2)$ and is given by the minimal energy
([11]):
\begin{displaymath}
d_{\tilde{g}}(F_1 , F_2)=(\int _{\gamma} G_{\tilde {g}}(F_1 (t),F_
1 (t))d\gamma)^{1/2}.
\end{displaymath}
Again, this metric structure $d_{\tilde{g}}$ should be adapted to the case of dualized Finsler structures. In 
particular, an associated Sasaki-type metric is also constructed
in a similar way.
\section{Quantum Mechanics versus Deterministic Hamilton-Randers Systems
} \label{comparison}

In this appendix, we compare the terminology and notions of deterministic finslerian systems  
with the respective notions of the usual Quantum Mechanics. Although not complete, the 
dictionary found suggests that we can understand all the basic notions of the 
Quantum Theory from deterministic finslerian theory. However, some
non-trivial differences appear. Hopefully, beyond the {\it more or
less} consistency of the ideas proposed in this work, these
differences will imply the possibility to test our proposal. The
differences are qualitative. Therefore, they can not confirm our
proposal but if they don't happen, Hamilton-Randers Deterministic Models
must be at least re-consider in another way or abandoned.

Form { Table 1} it is remarkable the following:
\begin{enumerate}
\item There is an ``inclusion" of the set of deterministic
Hamilton-Randers systems in the category of 
Quantum Systems. That means that we can describe deterministic
finslerian systems using Hilbert techniques as in usual quantum
systems. The problem at this level, is that there is not lower
bound of the ontological hamiltonian.

\item In any deterministic finslerian models there is an universal
minimal energy. Hopefully it could be related with the vacuum
energy or with dark energy density. Although the ontological
degrees of freedom refers to the Planck scale, the mass of the
particles associated with these degrees could be any other one.
Therefore one is free to associated the mass of the ontological
degrees of freedom to the cosmological constant.

\item The origin of the decoherence phenomenon in Quantum
Mechanics is  rather different in nature from our explanation of
the absence of interferences: in our case it is due to the
defining properties of the quantum system and the existence of
universal scales, associated with the structure of the vacuum and
the geometry of the phase space of the ontological degrees of
freedom.
\end{enumerate}

Because the existence of a Functor from the category of dynamical
systems to the category of deterministic finslerian models, we can
obtain a Quantum Mechanical description of the system. Using the
notion of two dimensional time, $U_t$-evolution and convex
invariance we can reconstruct systems resembling quantum systems
from deterministic finslerian models. We stress the possibility
that this map is surjective, that is, that all possible quantum
system can be obtained from deterministic systems, maybe through
the mechanism described here.

Deterministic finslerian models are not an usual hidden-variable
theory. There is not new interpretation for the wave function and
the meaning attached is the same as in the usual orthodox
interpretation of Quantum Mechanics: it describes the probability
amplitude to obtain a particular result when we measure the
properties of an individual quantum system. The main difference is
that we postulate deterministic systems for the degrees of freedom
at the Planck scale, from which quantum mechanics emerges. Quantum
Mechanics is then a complete theory at a phenomenological level.

\newpage
\begin{center}
{ Table 1: Deterministic Hamilton-Randers Systems/Quantum Mechanics}
\paragraph{}
\begin{tabular}{|r|l|}
\hline { Determ. Finsler. Systems} & { Quantum Mechanics} \\
\hline Maximal domain ${ K}_a$ & Quantum state $|a\rangle $ \\
\hline $U_t$ and $U_s$ evolutions & Quantum evolution $U_s$\\
\hline Coordinate invariant under $U_{t}$ & Beable Observable \\
\hline Coordinates not invariant with $U_t$ & Changeable Observable \\
\hline Selection of a point in ${ K}_a$ & Completion of the quantum state $|a\rangle $ \\
\hline Selection of a different & Different phase definition \\
Finsler metric $F$ & of the quantum states\\
\hline Convex invariance & Phase invariance of the quantum state \\
\hline Two deterministic dynamics &  Measurement process and evolution\\
                   & at the quantum  scale \\
\hline Existence of a minimal energy & Vacuum state\\
\hline ``Maximal Quantum Distance" $L_{max}=cT_max$ & Decoherence \\
\hline
\end{tabular}
\end{center}
\paragraph{}
\paragraph{}
{ Table 2: Differences between Deterministic Hamilton-Randers Systems
and Quantum Mechanics}
\begin{center}
\begin{tabular}{|r|l|}
\hline { Determ. Finsler. Systems } & { Quantum Mechanics} \\
\hline Maximal apparent speed & Unlimited apparent speed \\
for quantum correlations & for quantum correlations \\
\hline Maximal universal acceleration & Quantum maximal acceleration \\
 $A_{max}\sim 10^{52}\, m/s^2$ or $A_{max}\sim \frac{2m_{\nu} c^3 }{\hbar}$& $A_{max}\sim \frac{2m c^3 }{\hbar}$ \\
\hline The light is delayed due to & The speed of light is constant\\
the fluctuation of the geometry & \\
\hline Maximal coherence distance$\sim c/E_{min}$ & ? \\
\hline A small cosmological constant & A large cosmological constant \\
\hline Existence of a maximal eigenvalue for $\hat{{ H}}$& ?\\
\hline
\end{tabular}
\end{center}

The existence of a delay in the speed of light is also a
consequence of the relativity group in presence of maximal
acceleration. In this case, the form that maximal acceleration
deletes speed is given by the formula
\begin{equation}
\frac{dx}{ds}=c\sqrt{1-\frac{a^2}{a^2_{max}}}
\end{equation}
The underlying phenomenological geometry have locally a metric
with signature $(-1,1,1,1,1,1,1,1)$. Nevertheless, it is of
complete different effect than the delay in the quantum evolution,
that have at the end, an average constant speed.

We remark the significance for our scheme of the above predictions, than even 
qualitative, can falsify our approach. The first prediction is the main difference with quantum 
mechanics. We can not give a natural bound for the quantum correlations but if experiments are 
analyzed with enough precision and any trace of the bound for
quantum correlations is not obtained, our proposal should be
disregarded.

Some previous work was rather critic with the use of Finsler
geometry in Physics ([14]). Despite of it, a lot of research have
been done in the application of Finsler in field theory and
geometric dynamics ( for example, [15], [16] and references
there). Usually this work is beyond the constrains imposed in
[14], such that this negative result does not apply. Nevertheless,
our use of Finsler geometry, in particular Randers structures, is
with a very different purpose: to obtain an emergent Quantum
Mechanics, of the types briefly described for instance in [17].
Indeed, we need a kind of non-commutative description for the
fundamental degrees of freedom. In order to accomplish with
ergodicity in a finite time, the ontological degrees of freedom
are extended.


\newpage
\begin{thebibliography}{22}
\bibitem{the} Gerard't Hooft, {\it Determinism and Dissipation in Quantum gravity}, { 
hep-th/0003005}; Gerard't Hooft, {\it How does God play
dies?(Pre-) Determinism at the Planck Scale}, {
hep-th/0104219}.
\bibitem{the}R. Gallego, {\it A Hamilton-Randers version of 't Hooft Deterministic Quantum Models},
J. Math. Phys. { 47}, 072101 (2006).
\bibitem{the }R. Gallego, {\it A Riemannian structure associated to a Finsler structure}, { 
math.DG/0501058}.
\bibitem{the} D.Bao, S.S.Chern and Z.Shen, {\it An Introduction to Riemann-Finsler
Geometry}, Graduate Texts in Mathematics 200, Springer-Verlag.
\bibitem{the} D. Bao, Z. Shen,  {\it On the Volume of unit Spheres in a Finsler Manifold}, Results 
in Mathematics, { 26}(1994), 1.
\bibitem{the} Steven Weinberg, {\it The Quantum Theory of Fields, Volume I, Foundations}, 
Cambridge Monograph in Mathematical Physics.
\bibitem{the} Gerard't Hooft, {\it The mathematical basis for deterministic quantum mechanics
}, ITP-UU-06/14, SPIN-06/12,{ quant-ph/0604008}.
\bibitem{the} R. Gallego {\it On the Maximal Universal Acceleration in Deterministic Hamilton-Randers 
Models}, { gr-qc/0503094}.
\bibitem{the} E. R. Caianiello, {\it Quantum and Other Physics as System Theory}, La Rivista del 
Nuovo Cimento, { Vol 15, Nr. 4}(1992).
\bibitem{the} Gerard't Hooft, {\it The holographic principle}, { hep-th/0003004}.
\bibitem{the} R. Gallego, {\it Remarks on Determinisitic Quantum
Models based on Randers Spaces}, to appear.
\bibitem{the} M. Berger {A panoramic view of Riemannian Geometry}, Springer-Verlag (2002).
\bibitem{the} Olga Gil-Medrano, Peter W. Michor, {\it The Riemannian manifold of all Riemannian 
metrics}, { Quart. J. Math. Oxford. Ser (2) 42}(1991), 183.
\bibitem{the} J.D. Bekenstein, {\it The Relation between Physical and Gravitational Geometry}, 
Phys. Rev. { D48}, 3641(1993).
\bibitem{the} A. Bejancu, {\it Finsler Geometry and Applications}, Ellis Horwood Series in 
Mathematics and Applications, 1990.
\bibitem{the} S. Vacaru, P. Stavrinos, E. Gaburon and D. Gonta, {\it Clifford and 
Riemannian-Finsler Structures in Geometric Mechanics and Gravity}, Geometry Balkan Press, 
2005, { gr-qc/0508023}.
\bibitem{the} Stephen L. Adler, {\it Probability in Orthodox Quantum
Mechanics: Probability as a Postulate Versus Probability as an
Emergent Phenomenon}, { quant-ph/0004077}.
\bibitem{the} D. E. Kaplan and R. Sundrum, {\it A symmetry for the cosmological constant}, hep-th/0505265.
\bibitem{the} Florian Girelli, Stefano Liberati, Lorenzo Sindoni,
{\it Phenomenology of Quantum Gravity and Finsler Geometry},
gr-qc/0611024.
\bibitem{the} J.J. Sakurai, { Modern Quantum Mechanics} Addison-Wesley, 1994.
\bibitem{the}  Stephen L. Adler, {\it Statistical Dynamics of Global Unitary
Invariant Matrix Models as Pre-Quantum Mechanics}, hep-th/0206120.
\bibitem{the} Roger Penrose, {\it The Emperator's new mind}, Oxford University Press, New York 1989.
\bibitem{the} S. I. Vacaru, {\it Nonholonomic Ricci Flows: I. Riemann Metrics and Lagrange-Finsler Geometry
}, math.DG/0612162.
\bibitem{the} R.Erdem,{\it A symmetry for vanishing cosmological
constant in a extra dimensional toy model}, Phys. Lett. {
B621},11-17.
\bibitem{the} J. S. Bell, {\it Speakable and unspeakable in quantum
mechanics}, Cambridge University Press (1987).


\end{thebibliography}
\end{document}